\documentclass{svjour2}

\usepackage{amsmath,amsfonts,paralist}
\usepackage{amssymb}
\usepackage{xcolor}

\usepackage{graphicx}
\newcommand{\BEQ}{\begin{equation}}
\newcommand{\EEQ}{\end{equation}}
\newcommand{\BEA}{\begin{eqnarray}}
\newcommand{\EEA}{\end{eqnarray}}
\renewcommand{\H}{{\cal {H}}}

\definecolor{orange}{rgb}{0.9,0.7,0}

\definecolor{purple}{RGB}{153,50,204}

\begin{document}
\title{Critical study of hierarchical lattice renormalization group in magnetic ordered and quenched disordered systems: Ising and Blume-Emery-Griffiths models}

\author{F. Antenucci,$^{1,2}$ A. Crisanti$^{1,3}$ \\and  L. Leuzzi$^{1,2}$}
\institute{$^1$ Dipartimento di Fisica, Universit\`a {\em Sapienza},
  P.le Aldo Moro 2, I-00185 Roma, Italy. \\
  $^2$ IPCF-CNR, UOS Roma {\em Kerberos},  P.le Aldo Moro 2, I-00185 Roma, Italy.\\
  $^3$ ISC-CNR, UOS Sapienza,  P.le Aldo Moro 2, I-00185 Roma, Italy.}

\date{\today}

\maketitle

\begin{abstract}
Renormalization group on hierarchical lattices is often considered a valuable
tool to understand the critical behavior of more complicated statistical mechanical
models.  In presence of quenched disorder, however,  in many model cases predictions
obtained with the Migdal-Kadanoff bond removal approach fail to
quantitatively and qualitatively reproduce critical properties obtained in the mean-field
approximation or by numerical simulations in finite dimensions. 
In order to critically review this limitation
we analyze the behavior of Ising and
Blume-Emery-Griffiths models 
on more complicated hierarchical lattices.
We find that, apart from some exceptions,
the different behavior appears not only limited to Midgal-Kadanoff-like cells but is 
 associated right to the hierarchization of Bravais lattices in small cells also when in-cell loops are considered. 
\end{abstract}

\section{Introduction}

In this work we shall investigate the renormalization group analysis on spin 
systems with quenched disorder on hierarchical lattices. 
We will consider both Migdal-Kadanoff (MK) as well as more 
complex hierarchical lattices and we will study the critical behavior of systems with magnetic interactions  in presence of 
random fields and random exchange interactions.

Our main aim is to investigate whether hierarchical cells more complicated than MK ones and more similar to the local structure of 
short-range Bravais lattices can reproduce features of ferromagnets and spin-glasses 
so far unobserved in position space renormalization group studies with MK lattices. 
In order to obtain a more general comprehension  of the effect of bond moving 
we will provide estimates for critical quantities on several hierarchical lattices with different topology and  compare them to known
analytic and numerical results, when available.

Particular attention will be devoted to spin-glasses.
Understanding the nature of the low temperature phase of spin-glasses in finite
dimensional systems has turned out to be an extremely difficult
task. Since the resolution of its mean-field approximation, valid
above the upper critical dimension ($D=6$), more than thirty
years have passed without a final word about the possible
generalization of mean-field properties of spin-glasses to finite
dimensional cases.  The mean-field, else called Replica Symmetry
Breaking (RSB) theory \cite{Parisi79,Parisi80} involves a very interesting solution
for the spin-glass phase and its critical properties, rich of physical
(and mathematical) implications, and has been fundamental in solving
very diverse problems both in physics and in other disciplines \cite{Mezard87,Amit92,Mezard09}.
Because of its complicated structure, to overcome technical (maybe
also conceptual) obstacles hindering the ``portability" of RSB theory
predictions to short-range systems on Bravais lattice in $D<6$ is a
rather big challenge in theoretical physics.  Indeed, the RSB solution
is so complex that non-perturbative effects cannot be taken under
control in any perturbative loop-expansion around the upper critical
dimension and critical scaling behavior is yet to be understood
\cite{Chen77,DeDometal98,DeDominicis06,BraMoo11a,ParTem12a,SteNew12}. 
The main hindrance is the lack of
translational invariance in the position space for locally frustrated
systems with quenched disordered interactions, making the techniques
developed for quantum field theory and successfully exported to
statistical mechanical problems \cite{AmitBook,LeBellacBook} 
(e.g., for the Ising model critical
exponents)  inapplicable.

For what concerns { Kadanoff}  original approach in position space
\cite{Kadanoff, Ma76}, a proper extension of renormalization group techniques
to disordered and locally frustrated systems is still on its way. 
The generalization of classic position space renormalization methods on Bravais lattices
to disordered interaction, such as the ones
proposed for Ising spin models in the seventies \cite{Niemeijer73,Berker76},
has led to controversial results. On the one hand, by means of a
cumulant expansion approach, evidence for a spin-glass phase is yielded in
dimension two \cite{Kinzel78,Tatsumi78},  lower than the lower
critical dimension on the Bravais lattice: $D=2.5$ \cite{Franz94,Franz05b,Boettcher05}. 
On the other hand, the renormalization through block transformation on spin clusters does not yield
any spin-glass fixed point even in dimension three \cite{Kinzel78}.

The only  results  have been achieved using ``realizable'' approximations, 
namely those that are the exact solution of some alternative problem.
The first and most famous example is that of the classic ``bond-moving'' approximate Migdal-Kadanoff 
transformation, that when
applied to Ising models on Bravais lattices provides the exact solution for an Ising model on a very different 
lattice, \cite{Berker_79} known as ``hierarchical'' lattices \cite{KG_82}.
Note that, because of the ``bond-moving'' procedure, the hierarchical lattices corresponding to MK transformation
(MK lattices in the following) have basically a 1D topology, as, e.g.,  the ``necklace'' lattice in Fig \ref{necklace_MK}.

Therefore,  most of the position space renormalization group (PSRG) 
studies have been concentrating on hierarchical lattices for which, 
in the ordered cases, the renormalization group flow 
is indeed exact (no truncation required).
The study of these systems has brought to  important results, 
cf., e.g., Refs. \cite{Berker_79,Berker_82,Ohzeki08,Salmon10} and references therein. 

{However,  }MK lattices fail to represent short-range
spin-glasses on Bravais lattices also in the mean-field approximation
and are, thus, strongly limited in probing the actual nature of the
spin-glass phase \cite{Moore98,Ricci00}.
For what concerns perturbative disorder in ferromagnetic systems, it has been found that the Nishimori conjecture
  does not  provide the exact value of the multicritical point coordinate, \cite{Nobre01,Ohzeki08}
 although
recently Ohzeki, Nishimori and Berker \cite{Ohzeki08, Nishimori10} have put
forward an improved conjecture for models defined on hierarchical
lattices and they found that for various families of these models a
noteworthy recovery is obtained.

Furthermore the lack of translational invariance on hierarchical lattices
is supposed to make peculiarly difficult the study of first order transitions.
In particular, for pure fixed points while the expected first order transition is obtained
in some model (see for example \cite{Berker81,Ozcelik08}), there are relevant cases
in which the corresponding transition is missing on MK lattices \cite{Tsallis84,Tsallis_96}.
For what concerns disordered systems, 
MK lattices do not yield the first order fixed distribution
expected for the random Blume-Emery-Griffiths model \cite{Ozcelik08}, 
as obtained both in mean-field theory \cite{Crisanti05b} and by numerical simulations in finite dimension \cite{Paoluzzi10}.
We stress as in the latter case the MK lattices neither show the expected re-entrance in the phase diagram \cite{Ozcelik08,Crisanti05b,Paoluzzi10}.

Our aim is to test whether and which
of the above mentioned differences are specifically due to the bond
moving procedure at the basis of the renormalization group analysis. 
We will, thus, implement and compare the analysis of the critical behavior of well
known statistical mechanical models with quenched field and bond
randomness on both MK hierarchical lattices and the more complex ``folded 
hierarchical lattices", as we will call them. 

The latter family consists of hierachical lattices obtained applying the two root reduction directly to the 
Bravais lattice \cite{Tsallis_96} (see Figs. \ref{diamond}, \ref{WB_lattice}, 
\ref{Tsallis_2d_b3}, \ref{Tsallis_2d_b5}, \ref{Tsallis_3d}), 
without the bond moving specific of the Migdal-Kadanoff transformation.
So the final lattice has no longer just a 1D topology, but retains, in a small scale, the basic topology of the original lattice. 
So, unlike the MK family, in this case the original lattice is continuously reconstructed in the limit in which
the length of the basic cell, called $b$ in the following (see Sec. \ref{sec:Met}),
goes to infinity: the original lattice is a folded hierarchical lattice with an infinite basic cell.

In the following, we will  first critically revisit, in each model case, the analyses on hierarchical lattices  carried out in the  literature. 
We will, then, compare those results to the outcome of our studies on
{more complex lattices in the family of folded cells.}

The paper is organized as follows: 
{in Sec. \ref{sec:Ising}  we recall the implementation of the PSRG in the ferromagnetic Ising model. In  Sec. \ref{sec:Met}
we expose the details of the generalization of the PSRG in presence of generic quenched disorder.
In Sec. \ref{sec:RFIM}, we investigate the random field Ising model (RFIM) and in Secs. \ref{sec:Ising_2D} and 
\ref{sec:I3D} we report on the Ising spin-glass, respectively below and above the lower critical dimension.
We present
and compare the estimates of critical parameters and discuss how they comply to known
statistical mechanical  criteria in presence of disorder  (Nishimori conjecture, 
Harris criterion, ferromagnetic line inversion, $\ldots$). }
In Sec. \ref{sec:BEG3D} we consider the Blume-Emery-Griffiths model 
on several hierarchical lattices in dimension $d\geq 3$.
In the latter case our analysis shows a phase diagram displaying a 
reentrance for strong disorder, absent on MK 
lattices \cite{Ozcelik08}, but present in the mean-field approximation
\cite{Crisanti05} and in numerical simulations on 3D cubic lattices
\cite{Paoluzzi10,Leuzzi11,Paoluzzi11}.


\section{Hierarchical renormalization: Ising model}
\label{sec:Ising}
\begin{figure}[t!]
\begin{center}
\includegraphics[width=.55\textwidth]{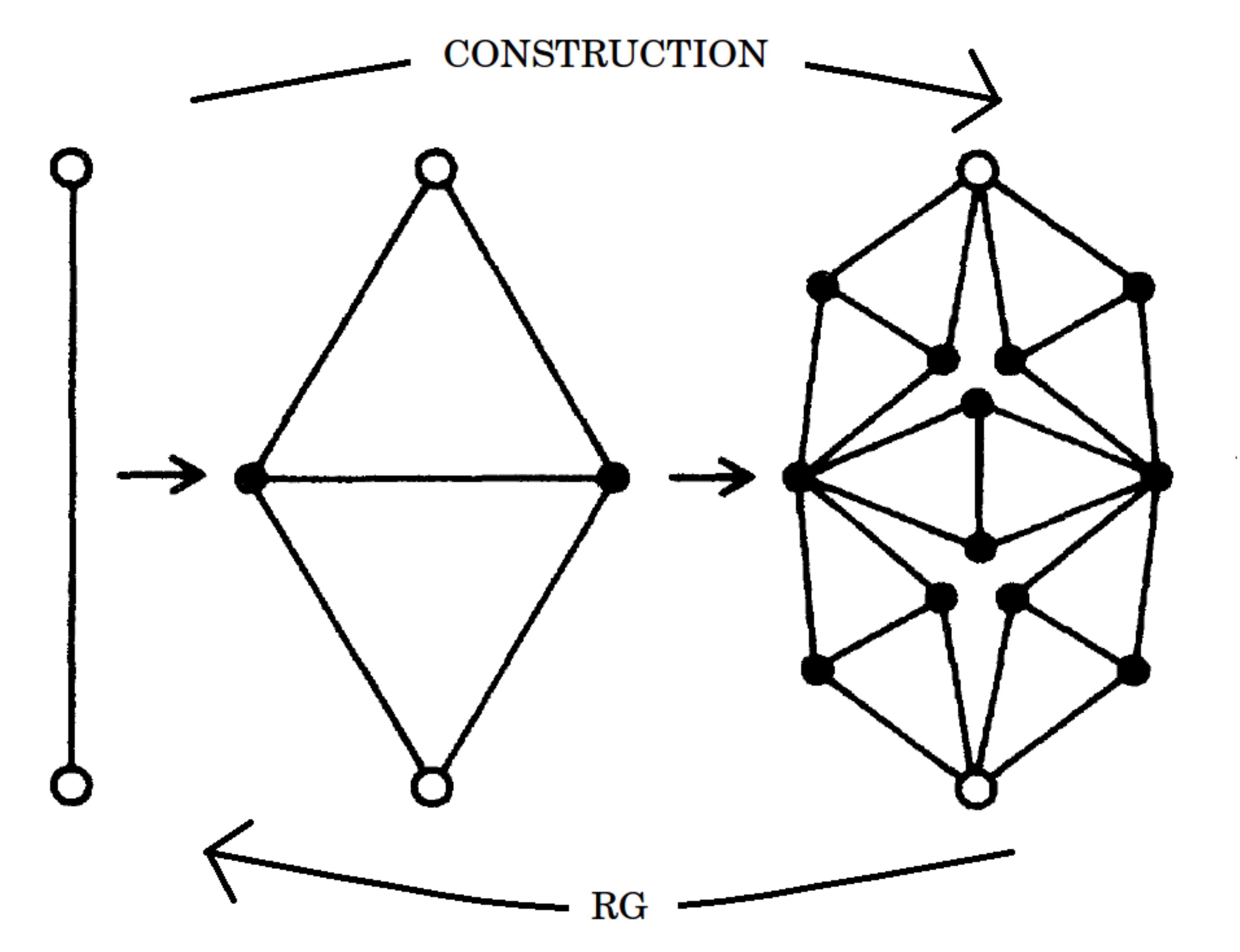}
\caption{Costruction of diamond, else called 2-dimensional Wheatstone-bridge, hierarchical lattice with fractal dimension $d=\log 5/\log 2= 2.3219\ldots$. }
\label{diamond}
\end{center}
\end{figure}

\begin{figure}[t!]
\begin{center}
\includegraphics[width=.12\textwidth]{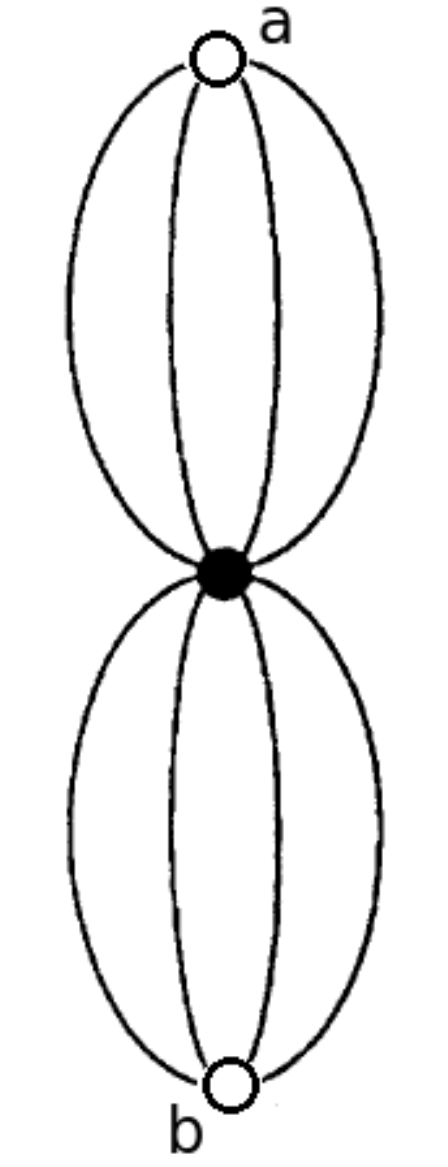}
\caption{{\em Necklace} MK lattice. It has $b=2$ and fractal dimension $d=3$.}
\label{necklace_MK}
\end{center}
\end{figure}

The Position Space Renormalization Group (PSRG) approach, approximated on
realistic Bravais lattices, becomes exact when  iterated on
Hierarchical Lattices (HL) \cite{Migdal,Kadanoff,Berker_79,KG_82,Tsallis_96}.
These lattices are constructed by carrying successive similar operations at each
hierarchical level.  E.g., at each level one replaces bonds by
well-defined unit cells. See, for example, Fig. \ref{diamond} 
for the diamond lattice or  Fig. \ref{necklace_MK} for a MK lattice.  
The PSRG procedure works the inverse way of the lattice generation, i.e.,
{{ one can implement it} }through a decimation of the internal sites of a given cell, leading to
renormalized quantities associated with the external sites.

In the pure case, the PSRG analysis proceeds as known by finding the
 interactions leaving the partition function invariant under decimation, and obtaining the critical exponents by the
eigenvalues of the first derivative matrix computed on the relative
fixed point \cite{Cardy_book}.

The well known Ising model is defined by the Hamiltonian
\begin{equation}
-  \H(s) = J \sum_{ \langle ij \rangle } s_i s_j + h\sum_i s_i,
 \label{f:Ham_Ising}
\end{equation}
where $s_i = \pm 1$ and $\langle ij \rangle$ indicates a sum over nearest-neighbor pairs. We stress that here and in the rest of the paper 
we include the temperature in the definition of the couplings (reduced parameters).

Since we, eventually, want to study the critical properties of disordered systems, 
and since through the renormalization group transformation original random bonds 
induce random fields (and vice-versa) already at the first step of renormalization,
it becomes more convenient to start using the following Hamiltonian
\begin{equation}
-  \mathcal{H}(s) = \sum_{\langle ij \rangle} \left[ J_{ij} s_i s_j + h_{ij} \frac{s_i+s_j}{2} + h_{ij}^{\dagger} \frac{s_i - s_j}{2} \right] 
\end{equation}
In this way each link between two sites $i$ and $j$ is associated
with three (possibly disordered) interactions $J_{ij}$, $h_{ij}$ and
$h_{ij}^{\dagger}$.

Decimating the inner sites $\{s\}$ of the basic cell ${\cal C}_{ab}$ of the
hierarchical lattice with external sites $s_a$ and $s_b$,  
while imposing the conservation of the  partition function of the cell
\begin{equation}
Z_{{\cal C}_{ab}}\equiv  x_{s_as_b}=\sum_{\{s\}\in {\cal C}_{ab}}\exp \left\{
- {\cal H}[s_a,s_b;\{s\}]
\right\},
\label{eq:partition}
\end{equation}
yields the renormalization group equations:
\begin{equation}\begin{split}
 J_R = \frac{1}{4} & \log \biggl( \frac{x_{++}x_{--}}{x_{+-}x_{-+}} \biggl) , \\
 h_R = & \frac{1}{2} \log \biggl( \frac{x_{++}}{x_{--}} \biggl) , \\
 h^{\dagger}_R = & \frac{1}{2} \log \biggl( \frac{x_{+-}}{x_{-+}} \biggl) , 
\end{split}
\label{eq_Ising}
\end{equation}
The partition sums $x_{s_a s_b}$, also called edge Boltzmann factors of the cell, 
are the weights of the cell for fixed external spins $s_{a}$ and $s_{b}$.
The sum in Eq. (\ref{eq:partition}) runs over all 
inner or free spins of the cell ${\cal C}_{ab}$. 

In the zero-temperature limit the relations become
\begin{eqnarray}
4 J_R\!\! &=&  \max\bigl[- \mathcal{H}(1,1,s)\bigr]
+ \max\bigl[- \mathcal{H}(-1,-1,s)\bigr]  \nonumber
 \\
 &-&\max\bigl[- \mathcal{H}(1,-1,s)\bigr]
 - \max\bigl[- \mathcal{H}(-1,1,s)\bigr] \nonumber 
\\
 2h_R \!\!&=&\max\bigl[- \mathcal{H}(1,1,s)\bigr]
 -\max\bigl[- \mathcal{H}(-1,-1,s)\bigr]\nonumber 
\\
 2h^{\dagger}_R \!\!&=&\max\bigl[- \mathcal{H}(1,-1,s)\bigr]
 -\max\bigl[- \mathcal{H}(-1,1,s)\bigr] \nonumber
 \\
\label{eq_Ising_zeroT}
\end{eqnarray}

When the external field is missing $h=h^{\dagger}=0$ and  $\mathcal{H}(s)=\mathcal{H}(-s)$, 
which implies $h_R=h^{\dagger}_R=0$. 

\subsection{The ordered ferromagnetic Ising model}
\label{ss:FMIM}

For an ordered ferromagnetic system ($J_{ij}=J$, $h_{ij}=h$ and $h_{ij}^\dagger=h^\dagger$)  the critical exponents can be obtained from the eigenvalues of the first derivatives matrix
\begin{equation} 
\left( \begin{array}{ccc}
\vspace{.3cm}
\frac{\partial J_R}{\partial J}  & \frac{\partial J_R}{\partial h} & \frac{\partial J_R}{\partial h^{\dagger}} \\
\vspace{.3cm}
\frac{\partial h_R}{\partial J} & \frac{\partial h_R}{\partial h} & \frac{\partial h_R}{\partial h^{\dagger}} \\
\frac{\partial h^{\dagger}_R}{\partial J} & \frac{\partial h^{\dagger}_R}{\partial h} & \frac{\partial h^{\dagger}_R}{\partial h^{\dagger}} \\
\end{array} \right)
\label{eq_JdJ}
\end{equation}
computed on the pure fixed point corresponding to the universality class of the ferromagnetic transition. 
The derivatives are easily obtained using
\begin{eqnarray}
 \frac{\partial x_{s_a s_b}}{\partial J}\!\! &=& \!\!\sum_{\{s\}\in {\cal C}_{ab}} \left(\! \sum_{\langle ij\rangle}s_is_j \!\right) \exp [-  \mathcal{H}(s_a, s_b,\{s\})] \, , \nonumber 
\\
 \frac{\partial x_{s_a s_b}}{\partial h}\!\! &=& \!\!\sum_{\{s\}\in {\cal C}_{ab}} \left(\! \sum_{\langle ij\rangle} \frac{s_i + s_j}{2} \!\right) \exp [- \mathcal{H}(s_a, s_b,\{s\})] \, , 
\nonumber 
\\
 \frac{\partial x_{s_a s_b}}{\partial h^{\dagger}}\!\! &=& \!\!\sum_{\{s\}\in {\cal C}_{ab}} \left(\! \sum_{\langle ij\rangle} \frac{s_i - s_j}{2} \!\right) \exp [- \mathcal{H}(s_a, s_b,\{s\})]. \nonumber
 \\
 \label{eq_der_BF}
\end{eqnarray}

\begin{figure}[t!]
\begin{center}
\includegraphics[width=.5\textwidth]{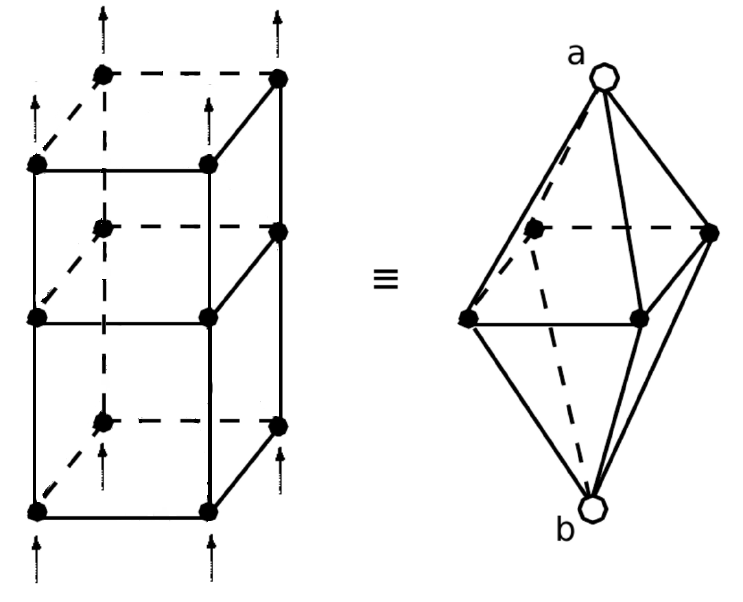}
\caption{Wheatstone bridge (WB) hierarchical lattice obtained  from a cubic lattice. It has $b=2$ and Hausdorff 
fractal dimension $d=\log 12 / \log 2 \approx 3.585$.}
\label{WB_lattice}
\end{center}
\end{figure}

\begin{figure}[t!]
\begin{center}
\includegraphics[width=0.8\columnwidth]{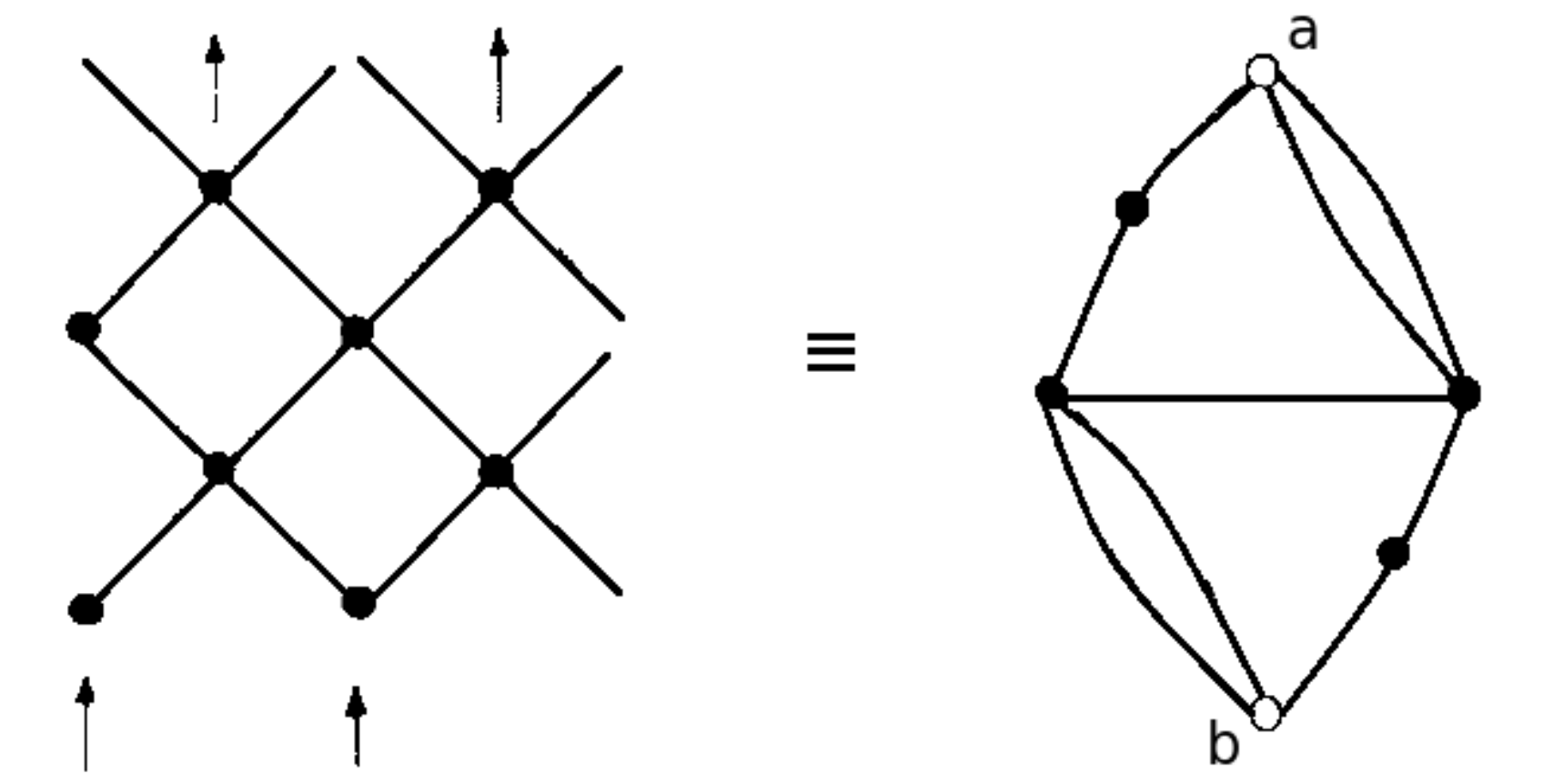}
\caption{Left hand side: square cells of cell spacing length $b=3$. 
  On the right side:  corresponding {\em folded square} hierarchical lattices with two  roots (open circles).  All  {\em outcoming}
  and {\em incoming} 
  sites, pointed by the arrows, are put together and
  generate, respectively, root sites {\tt a} and {\tt b}.  The
  inner sites of the square cell become the inner sites of the
  folded square hierarchical lattice. Note that with this construction we obtain self-dual 
  lattices with fractal dimension $d=\ln 9/\ln 3=2$. }
\label{Tsallis_2d_b3}
\end{center}
\end{figure}

\begin{figure}[t!]
\begin{center}
\includegraphics[width=0.9\columnwidth]{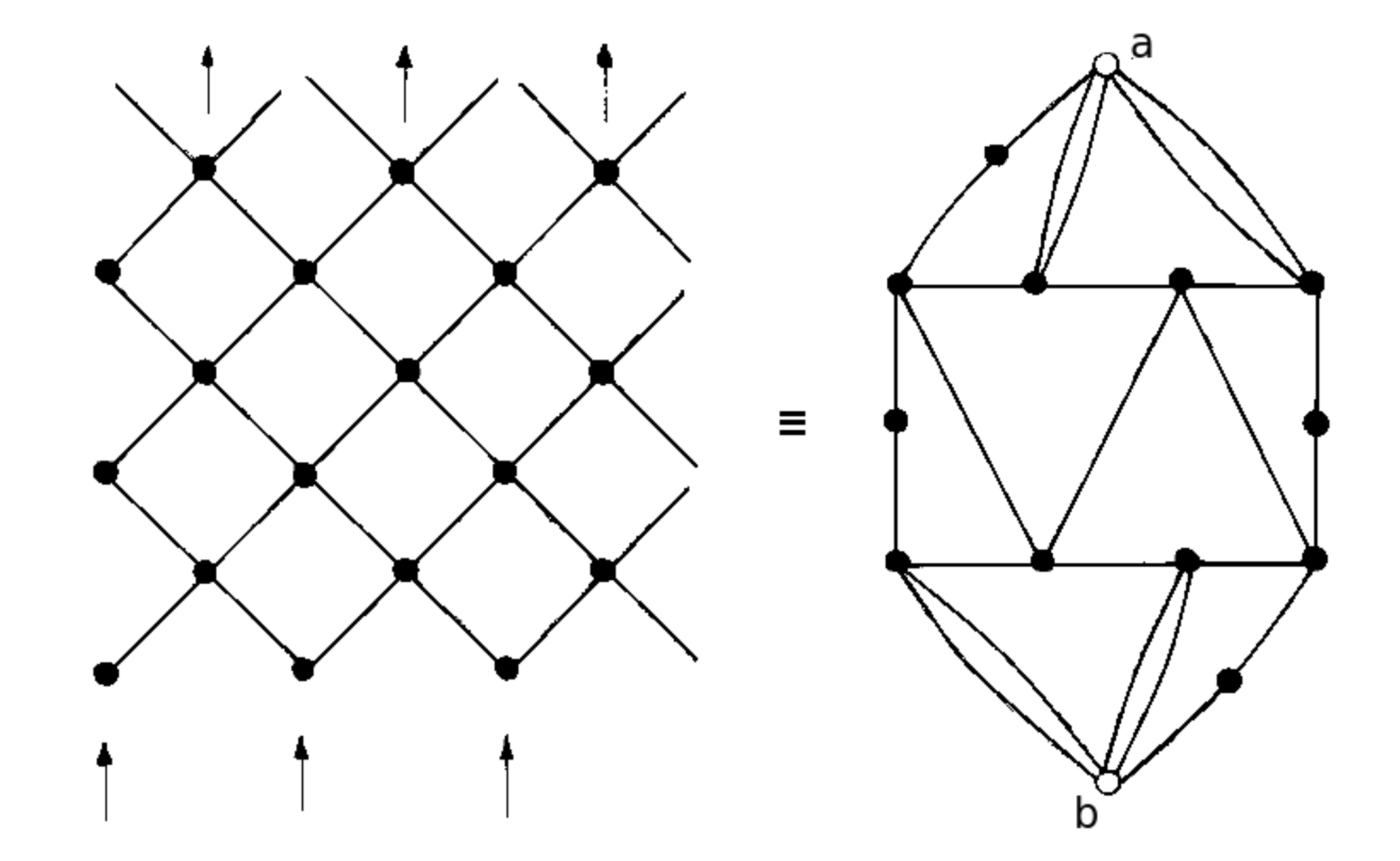}
\caption{Folded square lattice obtained as in Fig. \ref{Tsallis_2d_b3}, but for $b=5$.}
\label{Tsallis_2d_b5}
\end{center}
\end{figure}

In particular, if the fixed point is for $h=h^{\dagger}=0$, it is easy to see that the matrix in Eq. 
(\ref{eq_JdJ}) is diagonal and $\partial h^\dagger_R / \partial h^\dagger \equiv c$, 
where $c$ is the number of incoming (outgoing) links in the external outgoing (incoming) site. 
For example $c=4$ in Fig. \ref{WB_lattice}, $c=3$ in Fig. \ref{Tsallis_2d_b3} 
and $c=5$ in Fig. \ref{Tsallis_2d_b5}. 
In this case the only relevant eigenvalues are $\lambda_T = \partial_J J_R$
and $ \lambda_h = \partial_h h_R$, with the corresponding scaling exponents 
$y_{T,h}=\log_b \lambda_{T,h}$.

{ {The  critical scaling exponents of the physical observables are related to the scaling exponents $y_{T,h}$ by the scaling relations:}}
   \begin{align}
 \nu &= \frac{1}{y_T},
\\
 \eta &= d+2-2y_h \, ,
 \label{eq_Pure_Physical_Exponents_A}
 \\
  \alpha &= 2-\frac{d}{y_T}, 
\\
 \beta &= \frac{d-y_h}{y_T},
\\
 \gamma &= \frac{2y_h-d}{y_T},
\\
 \delta &= \frac{y_h}{d-y_h}.
\label{eq_Pure_Physical_Exponents_B}
\end{align}

The above PSRG scheme holds when the coupling constants $J_{ij}$ are all equal 
and the external field is ordered and homogeneous. 
To perform the equivalent analysis of the critical behavior in  disordered systems one has to generalize the 
 PSRG method to probability distributions of interaction parameters.

\section{{Hierarchical renormalization in presence of quenched disorder}}
\label{sec:Met}

In disordered systems the PSRG transformation is described by the
evolution of a probability distribution rather than single values of
coupling constants: \cite{Harris74b,Andelman84}
\begin{eqnarray}
P'({\cal J}_R)=\int \left[\prod_{\alpha=1}^{b^d}d{\cal J}_\alpha P({\cal J}_\alpha)
\right]
 \delta
\left[
{\cal J}_R - {\cal R}(\{{\cal J}\}_{b^d})
\right]
\nonumber
\\
\label{eq:P_J}
\end{eqnarray}
where ${\cal J}$ is the set of external parameters (couplings, fields,
chemical potentials, $\ldots$), $b$ is the  length of the cell in
lattice spacings (i.e., the scaling factor in the decimation procedure), $d$ the space dimension, 
so that $b^d$ is the size of the cell in number of bonds to be decimated, 
and ${\cal R}(\{{\cal J}\}_{b^d})$ is the local recursion relation for the interactions.

In MK lattices, because of their 1D-like topology, see, e.g.,  Fig. \ref{necklace_MK}, the transformation
can be divided into steps of so-called bond-moving and decimation, each of which
 involving only two bonds at a time.  
It is, thus, possible to exactly compute the probability distribution and represent it with
histograms, each bin of which characterized by a value of the
interactions and an associated probability \cite{Falicov_96}.


\begin{figure}[t!]
\begin{center}
\includegraphics[width=.99\textwidth]{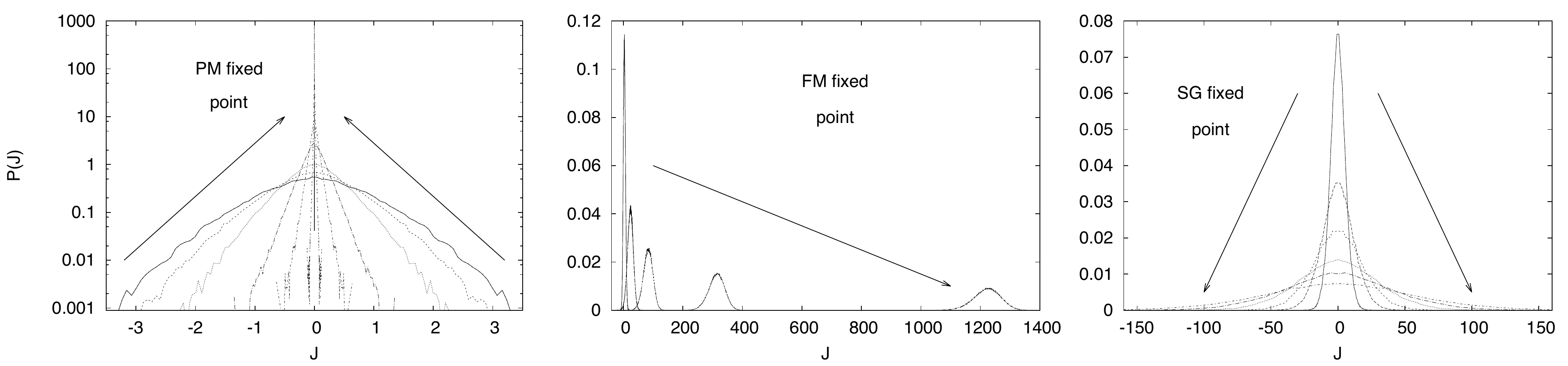}
\caption{RG evolution of the probability distribution of nearest-neighbor interaction 
                for the spin-glass Ising model on the Wheatstone bridge lattice of Fig. \protect{\ref{WB_lattice}} ($D\simeq 3.58$)
                and pool size $M=10^6$.
                Paramagnetic (left), Ferromagnetic (center) and Spin-Glasss (right) phases.
              }
\label{evo}
\end{center}
\end{figure}

 
In HLs built without the MK 1D-like structure, such as the Wheatstone-bridge-like in Fig. \ref{diamond},
this factorization is no longer possible and we must consider the
convolution of more than two links at a time. We are, eventually, obliged to
proceed in a statistical way.  The PSRG scheme is, then, accomplished by
representing the  probability distribution of the couplings by a pool of $M$ real
numbers \cite{Southern77}
from which one can compute its associated moments, at each
renormalization step. In the limit $M \rightarrow \infty$ these
moments should approach those of the exact renormalized probability
distribution.  
The process starts by creating a pool with $M$ coupling
constants generated according to the initial distribution.  
A PSRG iteration consists in $M$ operations in which one randomly picks
a set of $b^d$ couplings from the pool in order to generate one renormalized 
coupling, which will populate the renormalized pool. 
Following this procedure, one creates a new pool of size
$M$ representing the renormalized probability
distribution. During the PSRG procedure the moments of the coupling distribution
are of particular interest for the identification of the phases. 

For example, in Ising models with 
quenched disorder, 
 denoting by $J$ the average of the couplings 
and by 
 $\sigma_J$ the
mean square displacement, one obtains the Paramagnetic (PM), Ferromagnetic
(FM), and Spin Glass (SG) phases, as dominated by the
attractors
\vskip .1cm \qquad
\begin{tabular}{llr}
\vspace{1mm}
$ J \rightarrow  0;$ & $\sigma_J \rightarrow 0;$ & \quad PM ; \\
\vspace{1mm} 
$J \rightarrow \infty;$ & $\sigma_J \rightarrow \infty \; \; \left( J/ \sigma _J \rightarrow \infty \right); $ & FM ; \\
$ J \rightarrow 0;$ & $\sigma_J \rightarrow \infty;$ & SG . \\
\end{tabular}
\vskip .2cm

In Fig. \ref{evo} the typical PSRG iterations of the couplings distribution in the three phases are 
shown  for the random bond Ising model on the Wheatstone bridge lattice of  Fig. \ref{WB_lattice}. 

In order to reduce the dependence on a particular sequence of random
numbers, the evolution of each distribution is analyzed over $N_S$
different samples. 
This is especially relevant when the starting pool is near a critical point, and random
fluctuations can lead different samples of the same distribution into different attractors.
In this case we will adopt the convention that a phase is identified 
if at least $80\%$ of the $N_S$ samples flow into the same attractor. 
This defines the error for the location of critical points, which can be reduced by increasing the value of $M$.

{In presence of disorder it is hard to devise a  general prescription  for
finding the critical exponents, 
 like the one provided by  Eqs. (\ref{eq_JdJ}), (\ref{eq_der_BF}) for ordered models.
The idea is, then, to estimate the critical exponents 
by slightly perturbing the system from the  unstable fixed point distribution and 
measure how fast it departs from it under successive PSRG iterations.
We will discuss this procedure in detail in the following.}

\subsection{Random Field Ising Model}
\label{sec:RFIM}

In this section we discuss the PSRG study of the Random Field Ising Model (RFIM) with bimodal
and  Gaussian distributed quenched external field on the simple necklace 
MK  lattice of Fig. \ref{necklace_MK}, with fractal dimension $d=3$, and on the 
Wheatstone-Bridge (WB) hierarchical
lattice of Fig. \ref{WB_lattice}, with fractal dimension $d \approx 3.585$ \footnote{In a previous paper of Nobre and Salmon\cite{Nobre09}
the exact PSRG transformation of the hierarchical lattice is not achieved,
as pointed out by Berker \cite{Berker10}.}. 

The initial distribution of couplings for the RFIM reads
\begin{equation}
P(J_{ij},h_{ij},h^{\dagger}_{ij}) \!=\! \delta(J_{ij}-1) \, p(h_{ij})\, \delta(h^\dagger_{ij})
\end{equation}
where $p(h_{ij})$ is either a bimodal or a Gaussian  distribution:
\begin{equation}
 p(h_{ij}) \!=\! \left\{\begin{array}{l}
  \frac{1}{2}\left[\delta(h_{ij}\!-\!h_0)\!+\!\delta(h_{ij}\!+\!h_0)\right],
  \vspace{.3cm} 
 \\
 \frac{1}{\sqrt{2 \pi h_0^2}}  \exp
\left\{
-\frac{h_{ij}^2}{2 h_0^2}\right\}.
 \end{array}
 \right.
\end{equation}
The initial distribution is an even function of $h$, and $h^\dagger$.  
This symmetry is preserved under the PSRG transformation. 
To  maintain this symmetry in our finite sample, we, actually, use a pool of $2M$ interactions: 
for each of the $M$ computed renormalized interactions $(J_{ij},h_{ij},h^\dagger_{ij})$ we add to the pool
also the corresponding $(J_{ij},-h_{ij},-h^\dagger_{ij})$.

\begin{figure}[t!]
\begin{center}
\includegraphics[width=.99\textwidth]{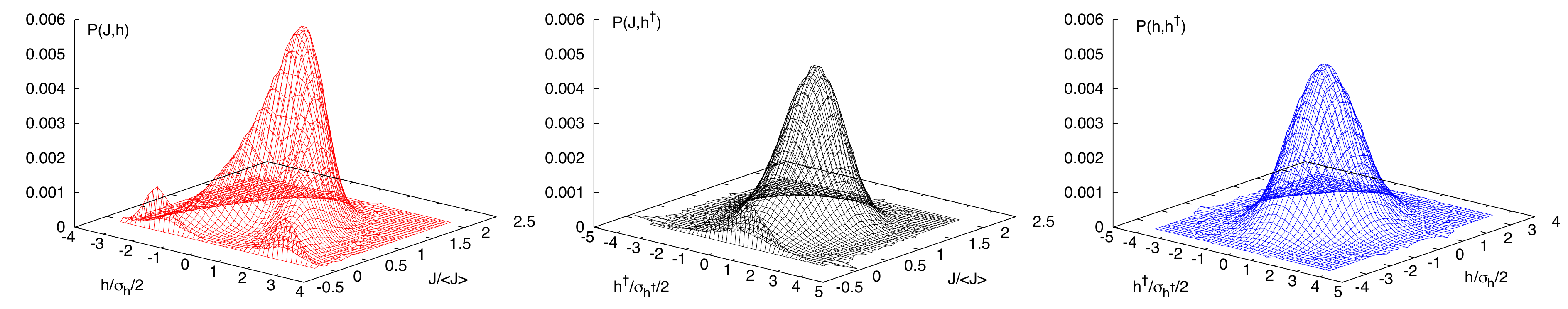}
\caption{
Projections of the critical $T=0$ fixed probability distribution for RFIM on WB $2$D lattice for bimodal initial distribution on $J,h$ (left), $J,h^\dagger$ (mid)
and $h,h^\dagger$ planes. }
\label{RFIM_fd_JH}
\end{center}
\end{figure}

This trick becomes important for long PSRG  iterations. As $M$ increases
the PSRG threshold step beyond which it becomes necessary quickly increases. This forced symmetrization is, thus, not crucial
in determining critical properties but helps decreasing finite size effects for small pools.
For the present computation we  take pools with up to $M=10^6$ interactions, enough to yield  statistically stable results.

The RFIM shows two phases identified by the behavior of the
ratio between the average coupling 
\begin{equation}
J=\langle J_{ij} \rangle
\end{equation}
 and the standard deviation of the fields 
\begin{equation}
h_0^2 = \langle h_{ij}^2 \rangle
\end{equation}
 along the PSRG flow: 
\vskip .1cm
\begin{tabular}{llr}
\vspace{1mm}
 $h_0 / J \to \infty$ & high temperature phase; \\
\vspace{1mm} 
$h_0 / J \to 0$ & low temperature phase;
\end{tabular}
\vskip .1cm
\noindent (and similarly for $h^\dagger$ field).
At the critical point both  $J$ and  $h_0$ (as well as 
$\sigma_{h^\dagger}$) flow to infinity, 
confirming that the critical behavior of RFIM is controlled by a zero-temperature fix 
point couplings distribution.

For the MK lattice of Fig. \ref{necklace_MK}  and for the WB lattice shown in Fig. \ref{WB_lattice}  the critical points 
are reported in Tab. \ref{tab:rfim_fp}.

\begin{table}[t!]
\begin{center}
\begin{tabular}{|l|c|c|}
\hline
Lattice & $h_0$ & $h_0/J$\\
\hline\hline
bimod. MK & $0.942(4)$ & $0.876(3)$\\
Gauss MK & $0.934(5)$  & $0.869(4)$\\
bimod. WB & $0.460(3)(2)$ & $0.443(5)$\\
Gauss WB & $0.456(3)$  & $0.448(4)$ 
\\ \hline
\end{tabular}
\caption{Critical fixed points exponents for RFIM on necklace MK lattice in Fig. \ref{necklace_MK} and WB lattice in Fig. \ref{WB_lattice}
with bimodal and Gaussian random field distributions.}
\label{tab:rfim_fp}
\end{center}
\end{table}


In  Fig.  \ref{RFIM_fd_JH} we show the projections of the critical fix point probability 
distribution for the bimodal case on the WB hierarchical lattice (in the Gaussian case 
they are qualitatively the same).

{At the zero-temperature fix point there are three independent exponents: the  scaling exponent of the distribution of couplings
$z$, the magnetic field scaling exponent $y_h$ and the thermal scaling exponent $y_T=1/\nu$ \cite{BM85}. 
They can be obtained following the  procedure proposed by Cao and Machta \cite{Cao92},
for both the bimodal and Gaussian case.}

{\em Distribution growth scaling.} $\qquad$
The exponent $z$ describes the deviation of the couplings probability distribution 
from the unstable fix point distribution under the PSRG transform. 
In order to estimate $z$, once the PSRG flux gets close to the 
fix point distribution, we fix the ratio $h_0 / J$ 
at the critical value in the subsequent PSRG iterations by
shifting at each step all the couplings $\{J_{ij}\}$ in the pool towards the ideal critical value.

{In particular, we compute the average $J$ in the given iteration and we compute the $J^*$ that the system should display if
$J^* =h_0/[h_0/J]_c$. Then we take  $J^*-J$ and we choose to change each one of the $J_{ij}$ in the distribution
 of the 80\% of  $J^*-J$. 
In our computation each one of these shifts turns out to be very small, of the order of 0.1\% for each $J_{ij}$. 
This is, though, an essential change 
because of the unstable nature of the fix point.}

We, then, evaluate the rescaling factor $\lambda\equiv h_0'/h_0 \simeq J/J'$
at each PSRG step next to the critical line and then take its average.
The exponent is estimated as
\begin{equation}
 z= \log_b \overline{\lambda} \, ,
\label{eq_z_exponent}
\end{equation}
where the overbar denotes the average over $10$ PSRG steps. Its values on the MK and WB cells both with bimodal and Gaussian distribution are shown in Tab. \ref{tab:RFIM}.
We note that   in all the cases it is $z \leq d/2$, that is the upper bound provided by Berker and 
McKay \cite{Berker86}.

{\em External field scaling.} $\qquad$
The exponent $y_h$ describes the rescaling of an infinitesimal homogeneous field and can be 
obtained by averaging the relations in Eq. (\ref{eq_JdJ}), (\ref{eq_der_BF})
over the fix point distribution
\begin{equation}
 y_h = \log_b \left< \frac{\partial h_R}{\partial h} \right> \, .
\end{equation}
Its values are reported in Tab. \ref{tab:RFIM} for the different cases. 
In all cases the value is smaller than the fractal dimension $d$, 
which is $3$ for MK and $\sim 3.585$ for WB, 
implying that the magnetization is continuous at the transition.

\noindent \indent {\em Correlation length scaling.} $\qquad$
In order to estimate the exponent $\nu$, we first reach a pool of renormalized couplings 
 satisfactorily representing the fix point distribution. 
Next we take a copy of the pool and generate a slightly perturbed couplings 
probability distribution by shifting every coupling 
$\{J_{ij}\}$ of the replicated pool by a small amount $\delta = 10^{-4} J$. 
The original and the perturbed pools are, then,  simultaneously renormalized. 
To reduce statistical fluctuations the couplings in the pool representing the fix point probability 
distribution are shifted, after each PSRG step, to keep the distribution close to the unstable fix point, 
in a manner similar to that used for the estimation of the exponent $z$.
Note that in this way the only role of the shift $\delta$ is to accelerate and make explicit the departure from the fixed point. 

By defining $t_n$ as the difference between the value of the ratio $h_0 / J$ 
in the two pools after $n$ PSRG iterations, the correlation length exponent is estimated as
\begin{equation}
 \frac{1}{\nu} = \log_b \overline{\left(\frac{t_{n+1}}{t_n}\right)} \, ,
\label{nu_tn}
\end{equation}
where the overbar denotes the average over the PSRG iterations $n$. 
Note that the argument of the logarithm is always positive, because 
leaving the fix point the second copy variance can either shrink or increase in its flux, but it does 
not oscillates between different PSRG steps. The sign of $t_n$ and $t_{n+1}$ is, thus, the same.

Typically we have $n= 3, \ldots,  9$, for which the perturbed pool is not too far from the unstable fix point, 
and the ratio $h_0 / J$  is  stable.
The result is  independent of $M$ and it is quite stable over independent PSRG evolutions, at least for $M\gtrsim10^5$.

We obtain  $1/\nu = 0.45 \pm 0.23$ for the bimodal distribution
and $ 1/\nu = 0.44 \pm 0.20 $ for the Gaussian distribution on the MK lattice.
Similarly, for the WB lattice we find
$1/\nu = 0.69 \pm 0.35$ for the bimodal distribution and $ 1/\nu = 0.68 \pm 0.31 $ 
for the gaussian distribution.
As summerized in Table ~\ref{tab:RFIM}.
Once the exponents $z$, $y_h$ and $\nu$ are known, the exponents $\alpha$ and $\beta$ 
are obtained from the scaling relations \cite{BM85}
\begin{eqnarray}
 \alpha &=& 2-(d-z) \nu,
 \label{alpha_z}
 \\
 \beta &=& (d-y_h) \nu.
 \end{eqnarray}

Notice that, at difference with Eq. (\ref{eq_Pure_Physical_Exponents_A}-\ref{eq_Pure_Physical_Exponents_B}), 
here the fixed point is at zero $T$ (as $1/J$) and the index $z \neq 0$.

The critical exponents obtained on MK lattice for the bimodal case are compatible with those 
obtained by Cao and Machta \cite{Cao92}. 
The exponents $y_h$ and $z$ depend strongly on the fractal dimension of the lattice, 
and their value for the $d=3$ MK lattice are in good agreement with the 
results for the three dimensional Bravais lattice, 
whilst for the $d\simeq 3.585$ WB lattice 
they are closer to that found for the four dimensional Bravais lattice, cf. Table~\ref{tab:RFIM}.

The value of the exponent $\nu$ is larger for the WB hierarchical lattice than for the MK lattice,
as in agreement with the behavior in hyper-cubic lattices of increasing dimensions.
Although the error bars for this exponent are large, we stress that in this case only the WB lattice gives
a numerical estimate compatible with the result for the cubic lattice. 

We, eventually, notice that the Gaussian and bimodal exponents are always compatible with each other and belong to the same universality class. 

\begin{table}[t!]
\begin{center}
\begin{tabular}{|l|c|c|c|}
\hline
Lattice & $z$  & $y_h$   & $1/\nu$    \\
\hline\hline
bimod. MK & $1.491(3)$ & $2.991(1)$ & $0.45(23)$   \\
{bimod. MK} \cite{Falicov95} & $1.4916(3)$ & $2.9911(2)$ & $0.445(2)$   \\
Gauss MK & $1.486(3)$  & $2.990(1)$ & $0.44(20)$   \\
bimod. WB & $1.788(2)$ & $3.575(1)$ & $0.69(35)$  \\
Gauss WB & $1.787(2)$  & $3.576(1)$ & $0.68(31)$ \\
3d MC \cite{Sim_RFIM} & $1.49(3)$ &  $2.988(4)$ & $0.73(5)$ \\
4d MC \cite{SIM_RFIM_4d} & $1.779(4)$ & $3.827(1)$ & $1.280(2)$ 
\\ \hline
\end{tabular}
\caption{Critical exponents at critical fixed distribution for RFIM on necklace MK lattice in Fig. \ref{necklace_MK} and WB lattice in Fig. \ref{WB_lattice}. 
The last two rows relate to simulations on Bravais 3D and 4D hyper-cubic lattices with bimodal distributed interactions.
{The more accurate exponents in Ref. \cite{Falicov95} are obtained using the histograms representation for the interactions probability distribution, that is
a feasible method only for MK cells.}
}
\label{tab:RFIM}
\end{center}
\end{table}

In the RFIM the bond are at first all equal, and the disorder is on site.
The major reason of moving from MK lattices to more complex HL is, actually,  the need of a better treatment of the bond structure.
In the next sections we move to models in which the disorder is on the bond from the outset.

\subsection{Random Bond Ising model in $d\leq 2.5$}
\label{sec:Ising_2D}

In this Section we consider the Ising model with bimodal $\pm J$ bond distribution on hierarchical lattices
mimicking the topology of the square lattice.

The initial probability distribution for interactions is 
\BEA
 P(J_{ij},h_{ij},h^{\dagger}_{ij}) &=& \left[ p \delta (J_{ij} - J) + (1-p) \delta (J_{ij} + J)\right] \,
\delta\left(h_{ij}\right)  \delta\left( h^\dagger_{ij}\right)
\label{eq:PpmJ}
\EEA 
where $J>0$ and $p\in[0,1]$ is the probability of a ferromagnetic bond.

For low enough $p$, this model on regular lattice has an antiferromagnetic phase.
On hierarchical lattices the antiferromagnetic order is preserved under PSRG only when the rescaling factor $b$ is odd,
so that a symmetric phase diagram in $p \leftrightarrow (1-p)$ is obtained, where the ferromagnetic phase is 
replaced by the antiferromagnetic phase and vice-versa.

To capture some features of Bravais lattices, we shall focus on 
hierarchical lattices with elementary cell more complex than the $1$D-like MK cells.  
In particular we shall consider the cell proposed
by Nobre \cite{Nobre01}, Fig.  \ref{Tsallis_2d_b3}, with rescaling factor $b=3$ and its
extension to $b=5$, Fig.  \ref{Tsallis_2d_b5}.

{\em Nishimori conjecture.}\qquad 
Even though the Random Bond Ising model in 2D does not display any spin-glass phase \cite{Ohzeki09},
this model on hierarchical lattices 
is an excellent play ground to test 
Nishimori's conjecture \cite{Nishimori01, Nishimori02}. 
Here we briefly recall what the conjecture is.
 
The idea behind it stems from noting that 
the partition function $\mathcal{Z}$ of non-random Ising models on self-dual lattices
is itself self-dual \cite{Wegner}.

Let us express $\mathcal{Z}$ in terms of the edge Boltzmann factors $u_{\pm 1} (J) = e^{\pm J}$
(note the use of the reduced parameters). The Fourier transform of $u_{\pm 1} (J)$, 
$u^*_{\pm 1} (J)$,  is the dual Boltzmann factor \cite{Wu_Wang}.
As a consequence, the critical point of a self-dual model is obtained 
by the fix point condition $u_{\pm 1} (J) = u_{\pm 1}^* (J) $, 
which yields $J_c =\frac{1}{2} \log (\sqrt{2}+1)$.

Within the replica method approach,  the 
relevant variables 
are the averaged edge Boltzmann factors  $x_k(p,J)$, 
which correspond to the configuration with the spin connected by the bond equal to +1 in $n-k$ 
replicas and $-1$ in remaining $k$ replicas. 
Self-duality is now expressed 
by the invariance of $\mathcal{Z}_n \equiv \overline{\mathcal{Z}^n}$ under the 
simultaneous exchange $x_k(p,J) \leftrightarrow x^*_k(p,J)$ for all $k$.
The overbar denotes the average over quenched bond disorder. 
Unlike the non-random case, it is not possible
to identify the  critical point from the fix point condition 
of the duality relations, because the relations 
with $k=0,\ldots, n$ are not satisfied simultaneously. 
The Nishimori's conjecture \cite{Maillard03,Takeda}, then,  identifies a 
point $(p_N,J_N)$, {{ called "multicritical", by means of a}} fix point condition for the leading $k=0$ Boltzmann factor
\begin{equation}
 x_0(p_N,J_N) =x^*_0(p_N,J_N) \, ,
\label{eq:Nishimori_c1}
\end{equation}
on the Nishimori line $e^{-2J}=(1-p)/p$,\cite{Nishimori81,Nishimori01} where enhanced symmetry simplifies the system 
properties significantly. { {This point of the Nishimori line is called multicritical because, 
when a SG phase is present, it is the point at which PM, FM and SG phases are all in contact with each other. This does not occurs
in 2D, see, e.g., Fig. \ref{fig_Tsallis_2d} but it does occur in 3D, cf. Fig. \ref{P-D_Tsallis_50k}.}}

The conjecture is proved exact for $n=1,2$ and $\infty$. 
In the limit $n \to 0$ the condition $x_0 = x^*_0$ on the Nishimori line becomes
\begin{equation}
H(p_N)\equiv -p_N \log_2 (p_N) -(1-p_N) \log_2(1-p_N) = \frac{1}{2} \, ,
\label{eq:Nishimori_conjecture}
\end{equation}
where the function $H(p_N)$  is the binary entropy.

This conjecture turns out to be  wrong for some systems on HL.\cite{Hinczewski05,Ohzeki08}
Ohzeki, Nishimori and Berker \cite{Ohzeki08,Nishimori10} noted that for HL's a systematic 
approximation for the multicritical point can be obtained 
by imposing  Eq. (\ref{eq:Nishimori_c1}) at each PSRG transformations.

Note that in the two-dimensional case, even though the SG
phase is absent, the Nishimori point is expected to coincide
with a critical point, unstable along the phase boundary;
so, as it is usually done, in the following we identify the Nishimori point as the 
intersection between the Nishimori and the critical lines.

Here we test the original Nishimori conjecture on a Folded Square (FS) hierarchical lattice
constructed through a bond-moving procedure that, at difference with the
MK bond-moving prescription, retains the local correlation of the bonds, 
see Figs. \ref{Tsallis_2d_b3} and \ref{Tsallis_2d_b5}.


\subsubsection{ Numerical results}
\label{Sec:Num_res_RB_2d}
\begin{figure}
\begin{center}
\includegraphics[width=\columnwidth]{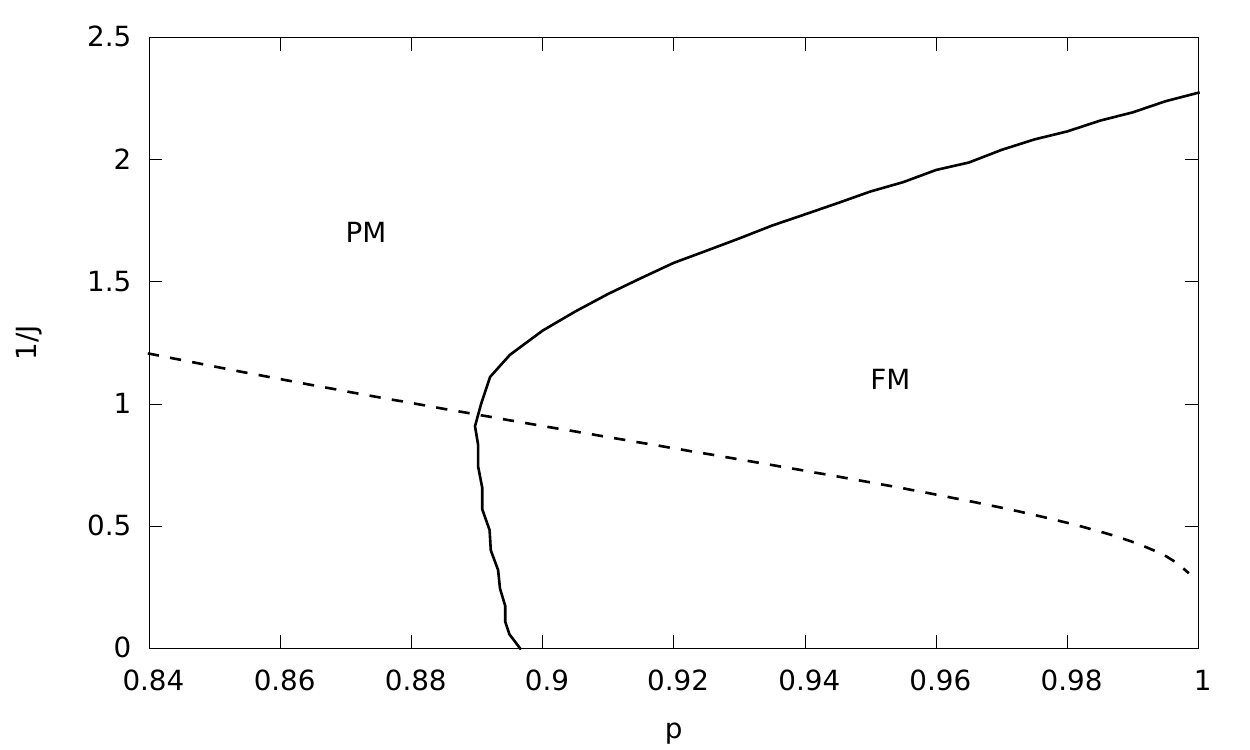}
\caption{Phase diagram of Ising 2D with ``folded square" HL with $b=5$.  The
  diagram is symmetric in $p \rightarrow 1-p$ and the
  anti-ferromagnetic part is not shown. The dashed line represents the Nishimori line. 
  }
\label{fig_Tsallis_2d}
\end{center}
\end{figure}

In Fig. \ref{fig_Tsallis_2d} we show the $p,T(=1/J)$ phase diagram obtained
from the PSRG analysis in two dimensions on the  lattice
shown in Fig. \ref{Tsallis_2d_b5} with rescaling factor $b=5$.

For $p=1$ we find the critical temperature is $T_c=2.269185(1)$, 
in agreement with the Onsager solution $T_c=2/\log(1+\sqrt{2}) \simeq 2.269185$. 
This result, also found for the folded square lattice of Fig. \ref{Tsallis_2d_b3} with
$b=3$,\cite{Nobre01} and WB with $d=2.32$, \cite{Salmon10}  
follows from the duality properties of the unit cells.\cite{Tsallis_96}

{\em Nishimori point.} \qquad
The position of the multi-critical point expected from Nishimori's conjecture,  
Eq. (\ref{eq:Nishimori_conjecture}), is $p_N \simeq 0.889972$.
For the cell of Fig. \ref{Tsallis_2d_b3} the estimated  value is $p_N = 0.8903(2)$
$(2H(p_N)=0.998(1))$\cite{Ohzeki08} with $T_N=0.9557(18)$.\cite{Nobre01}

For the cell of Fig. \ref{Tsallis_2d_b5}, performing $N_S=20$ independent PSRG 
calculations with a poll of $M=10^5$ initial bond configurations, we 
estimate (cf. Table  \ref{tab_multicritical_2d})
\begin{eqnarray}
 p_N &=& 0.8902(1), \quad T_N = 0.9571(1)  
 \nonumber \\
 2H(p_N)&=&0.9985(1) \, ; \nonumber
\end{eqnarray}
yielding no difference with respect to  the cell with $b=3$.
We conclude that the conjecture fails also on this more complex
hierarchical lattice.

The folded square cell estimates with $b=3$ or $b=5$ turn out to be in good agreement with 
each other and with
the estimate $p_N = 0.8905(5)$ given by the transfer matrix approach, \cite{Reis99} but are
slightly larger than those from high temperature series expansion,  
$p_N = 0.886(3)$, \cite{Singh96} and  Monte Carlo simulation, $p_N = 0.8872(8)$. \cite{Ozeki98}

In Table \ref{tab_multicritical_2d} we report the results on critical points for an easier comparison.
 
{\em Critical slope. $\qquad$ }{A  quantity usually studied} is the slope of the critical line
close to $p=1$:\cite{Harris74}
\begin{equation}
 s\equiv {\frac{1}{T_c(1)}} { {\frac{dT_c(p)}{dp}} \bigg| _{p=1}} \, . 
\label{slope}
\end{equation}
The Domany's perturbative approach \cite{Domany79} yields $s=2\sqrt2 / [\ln (\sqrt2 + 1) ] \simeq 3.209$
for the Ising with $\pm J$ bond distribution on square lattice. 
The approach assumes {\em weak} disorder, i.e.,  a
qualitatively irrelevant disorder that does not undermine the existence of the ferromagnetic phase at
low $T$ and does not change the universality class of the PM/FM transition for $p<1$. 
By computing $s$ it is, thus, argued  that one can
discriminate whether quenched disorder is a relevant perturbation,
causing a change in the universality class. \cite{Ohzeki11}
Ohzeki and collaborators \cite{Ohzeki11} suggest that Domany's method can be applied to any self-dual lattice, and then 
one can probe the relevance of disorder from the slope $s$ also for the HL 
of Figs. \ref{Tsallis_2d_b3} and \ref{Tsallis_2d_b5}. 

The duality approach gives $s=3.27866...$ \cite{Ohzeki11} 
for the lattice in Fig. \ref{Tsallis_2d_b3} with $b=3$.
{From a best fit of the points close to $p=1$ with a  pool of size $M=5 \cdot 10^6$ with $N_S = 20$ samples  we obtain
 $s=3.30(3)$, compatible with, though less precise than,  the duality estimate.
In the $b=5$ case of Fig. \ref{Tsallis_2d_b5},   
with a  pool of size $M=5\cdot 10^6$ with $N_S = 20$ samples we find $s=3.32(3)$.
Both values are different from Domany's prediction $s \simeq 3.209$.\cite{Domany79} 
This is not, however, a strong issue in favor of strong disorder.}

\begin{figure}
\begin{center}
\includegraphics[width=\columnwidth]{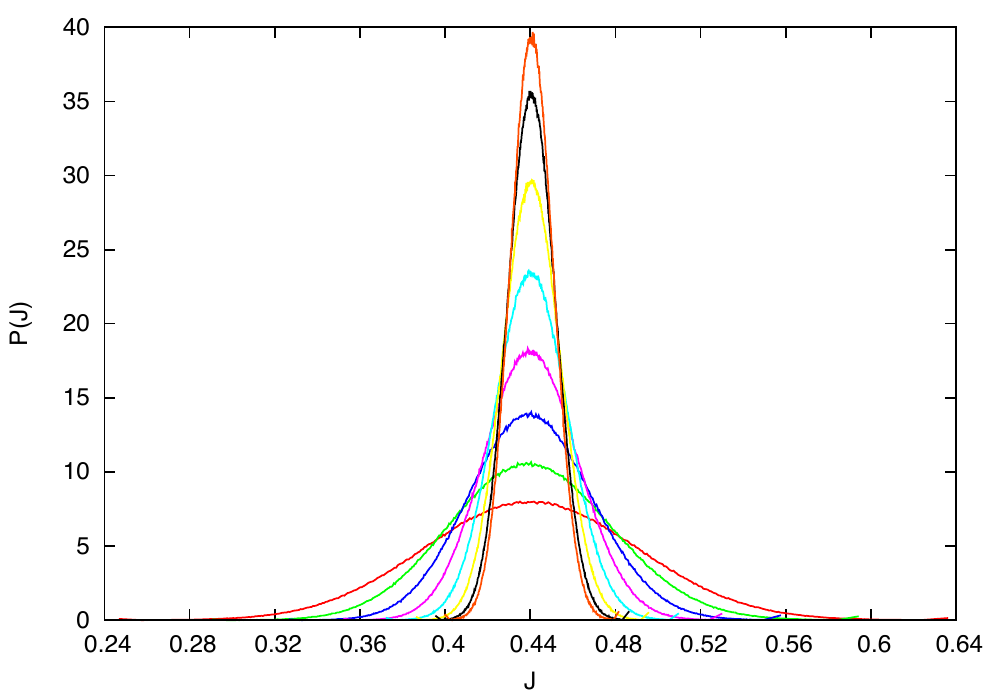}
\caption{ PSRG evolution of the coupling distribution on the critical line near the pure fixed point (steps from 4 to 11 are shown for $p=0.01$): 
  the width quickly decreases, so that the distribution tends to a Dirac delta and the disorder disappears.
  No other fixed distribution. besides the pure criticality, are obtained on the PM-FM transition line.
  }
\label{fig_Tsallis_c-evo}
\end{center}
\end{figure}

{According to the argument of Ref. \cite{Ohzeki11} this would imply that the disorder
should be relevant in these cases. 
However, if it is true that a slope equal to Domany's slope might be consistent with the
hypothesis of irrelevant disorder, the fact that the slope is different from Domany's value  does not, actually, imply the opposite (i.e., the existence of a new "strong disorder" fixed point). 
Indeed, the HL RG approach already fails to quantitatively precisely capture the critical behavior of the system and the lack of a numerical coincidence is just a consequence of this.
One can confirm this by direct inspection, yielding no other fix points
after disorder is introduced and, consequently, no universality class change.}
Any coupling distribution on the transition line tends to become more and more 
peaked under PSRG transformations, 
tending to the pure FM-PM critical fixed point, as shown in Fig. \ref{fig_Tsallis_c-evo}. 
Disorder in the two dimensional Ising model does not appear to play any role 
on any lattice cell analyzed here (see also the Harris criterion below).

\begin{table}
\begin{center}
\begin{tabular}{|c|c|c|c|}
\hline
       HL / Method & Fig./Ref.   & $p_N$& $p_{T=0}$\\
\hline
FS $b=3$ & \ref{Tsallis_2d_b3}/\cite{Nobre01} & $0.8903(2)$ & $0.8951(3)$\\
FS $b=5$ & \ref{Tsallis_2d_b5} & $0.8902(1)$ & $0.8966(2)$\\
\hline
& \cite{Reis99}& $0.8905(5)$ &\\
Square &\cite{Singh96}& $0.886(3)$&\\
& \cite{Ozeki98} & $0.8872(8)$ &\\
&\cite{Kawashima97} & & $0.896(1)$\\
&\cite{Kawashima97} && $0.894(2)$ \\
\hline
Nishimori conjecture & \cite{Nishimori01} & $0.889972...$& \\
\hline 
\end{tabular}
\caption{Multicritical point as computed on different HLs and different estimates on the square lattice.}
\label{tab_multicritical_2d}
\end{center}
\end{table}


{\em Harris criterion.}\qquad 
The widely accepted form of the Harris criterion \cite{Harris74, Chayes} is that in
ferromagnetic systems with random interactions the randomness is irrelevant if $\alpha$, the specific 
heat exponent of the
corresponding pure system, is negative, while for systems with positive $\alpha$ the random system 
exhibits different critical behavior in presence of disorder. 
For the folded square lattices for both $b=3$ and $b=5$ we find $\alpha<0$.

Though it is known that this criterion can fail on 
HL if the bonds in the rescaling volume are not all equivalent,\cite{Andelman84,Kinzel81,Andelman85,Derrida,Sutapa,Efrat} 
in the present case the Harris criterion is, actually, satisfied since the disorder is irrelevant, as discussed above with respect to the critical slope value.

{\em FM line reentrance.} \qquad
An important feature of the $p,T$ diagram, according to the duality
requirements, is the reentrance of the transition line below the
multicritical point: $p_N > p_{T=0}$.  
 The zero temperature transition point $p_{T=0}$ can be estimated by
finite size scaling analyses 
of the ground state.
Calling $E_p$ and $E_a$ the ground-state energies with, respectively, periodic and  
anti-periodic boundary conditions in one direction, 
and $\Delta=E_p-E_a$ the domain wall energy, 
we can determine  two estimates of critical concentrations of antiferromagnetic bonds by
looking at  the point where  the asymptotic  dependences $[\Delta]$ and $[\Delta^2]^{1/2}$ change
from increasing to decreasing, where the average $[\ldots ]$ is taken on different bond samples.
More explicitly, by defining 
\begin{equation}
 [\Delta] \sim L^\rho \quad \text{and} \quad [\Delta^2]^{1/2} \sim L^{\theta} \, , 
\end{equation}
via the determination of exact ground states for large system sizes and huge
sample numbers, Kawashima and Rieger give 
the estimates $p_{T=0}^{(\rho)}=0.896(1)$ and $p_{T=0}^{(\theta)}=0.894(2)$ 
looking at the point where  $\rho$ and $\theta$  {{ respectively}}
change sign \cite{Kawashima97}.
The PSRG approach with the folded square of Fig. \ref{Tsallis_2d_b3}  ($b=3$) leads to
$p_{T=0}=0.8951(3)$ \cite{Nobre01}, 
while with the cell of Fig. \ref{Tsallis_2d_b5} ($b=5$) we find  $p_{T=0}=0.8966(2)$
as summarized in Tab. \ref{tab:RFIM}.

The critical indices for the pure ferromagnetic point can be evaluated as discussed in Sec.~\ref{ss:FMIM}:
we obtain  $y_T = 0.7303(1)$ and $ y_h = 0.8518(1)$ for the folded square lattice with $b=3$ and  for the folded square lattice with $b=5$ we find $y_T =0.7589(1)$ and $ y_h=1.059(1)$.
In both cases they are not consistent with the values of Onsager solution,
which read $y_T = 1 $ and $y_h=1.875 $, respectively. 
From the knowledge of $y_T$ and $y_h$ the critical indexes are obtained from the usual scaling relations, cf. Eq. 
(\ref{eq_Pure_Physical_Exponents_B})
and their numerical values are reported and compared on Table \ref{tab_ce_2d_tot}.


{\em Zero temperature stiffness.}\qquad
We conclude this section by discussing the exponent $\nu$ of the zero-temperature 
spin-glass transition for the case of a Gaussian
distribution of bonds with zero mean and initial width $\sigma_J$.
It can be obtained directly from scaling of $\sigma_J$ under PSRG : 
\begin{equation}
 \sigma'_J (b) \sim \sigma_J b^\theta .
\label{eq:scaling_theta}
\end{equation}
The sign of the stiffness exponent $\theta$ is directly related to the low temperature
phase: for positive (negative) $\theta$ the system scales under PSRG flow
towards strong (weak) couplings, distinctive of a low temperature
spin-glass (high $T$ paramagnetic) phase. 
For continuous and symmetric probability distributions $P(J)$, 
the temperature $T$ appears in the PSRG equations as a dimensionless ratio between couplings,
so that the scaling (\ref{eq:scaling_theta})
is equivalent to $T \sim L^\theta $, or $L \sim T^{1/\theta}$. 
In a phase transition at $T \rightarrow 0$ the latter scaling can be identified 
with the scaling of the correlation length $\xi \sim T^{-\nu}$ implying \cite{Bray84}:
\begin{equation}
 \nu = - \frac{1}{\theta} .
\end{equation}
\begin{table}
\begin{center}
\begin{tabular}{|l|c|c|c|}
\hline
         & FS b=3 & FS b=5 & Exact \\
\hline 
$\alpha$ & -0.7385(1) & -0.6353(1) & 0 \\
$\beta$  & 1.572(1) & 1.240(1) & 0.125 \\
$\gamma$ & -0.4057(1) & 0.1558(1) & 1.75 \\
$\delta$ & 0.7419(1) & 1.126(1) & 15 \\
$\nu$    & 1.369(1) & 1.318(1) & 1 \\
$\eta$   & 2.296(1) & 1.882(1) & 0.25 \\
\hline
\end{tabular}
\caption{Critical indices of pure ferromagnetic critical point for Ising model on the folded square 
      lattice for $b=3$, Fig. \ref{Tsallis_2d_b3}, and for $b=5$, Fig. \ref{Tsallis_2d_b5}. }
\label{tab_ce_2d_tot}
\end{center}
\end{table}

\begin{table}
\begin{center}
\begin{tabular}{|c|cc|c|}
\hline
       Lattice   type & Fig. & Ref.   & $\theta$ \\
       \hline
MK $b=2$ &&\cite{Southern77}]& $-0.270(2)$\\
MK $b=3$ & & \cite{Nobre98}&$-0.278(2)$\\
WB $b=2$       &\ref{diamond}  & \cite{Tsallis_96}  &$-0.290(3)$\\
WB $b=3$     &  &\cite{Nobre98}  &$-0.298(2)$\\
FS $b=3$       &  \ref{Tsallis_2d_b3} & \cite{Tsallis_96} &$-0.275(1)$\\
FS $b=5$       & \ref{Tsallis_2d_b5} &&$-0.2714(2)$\\
\hline
&&\cite{Bray87} & $-0.291(2)$ \\
Square &&\cite{Bray02} & $-0.287(4)$\\
& &\cite{Weigel07} & $-0.284(4)$ \\
\hline
\end{tabular}
\caption{Stiffness exponent $\theta$ on different HLs and different estimates on the square lattice.}
\label{tab_stiffness_2d}
\end{center}
\end{table}

\begin{figure}
\begin{center}
\includegraphics[width=0.93\columnwidth]{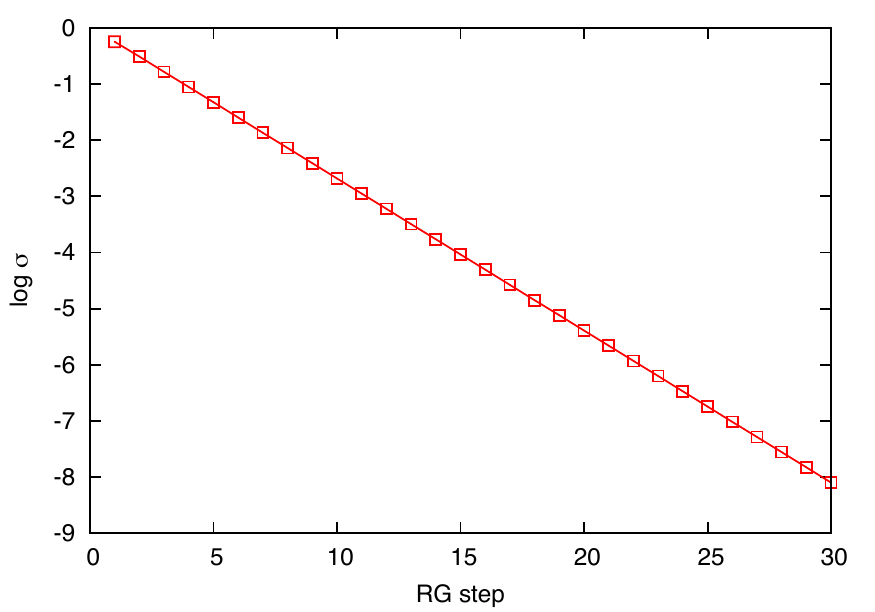}
\caption{Standard deviation of the Gaussian probability distribution
  of quenched disorder at $T=0$ vs PSRG iteration steps .}
\label{logJ_vs_n_G}
\end{center}
\end{figure}

In Fig. \ref{logJ_vs_n_G} the behavior of $\sigma_J$  is shown as function of PSRG steps   for the 
case of zero-average Gaussian initial bond distribution on the folded square cell of Fig. 
\ref{Tsallis_2d_b5}.
From this we get  $\theta=-0.2714(2)$,  leading to $\nu = 3.685(3)$.
For the cell of Fig. \ref{Tsallis_2d_b3}, with $b=3$ it was found $\theta=-0.275(1)$
\cite{Nobre98}. 
We report in Table  \ref{tab_stiffness_2d} the values obtained on different HLs \cite{Southern77,Tsallis_96,Nobre98,Salmon10} 
and on the square Bravais lattice \cite{Bray87,Bray02,Weigel07}.
For the folded square with both $b=3$ or $b=5$ the value is similar to that found for the MK lattice, 
whilst for the WB lattices the values are closer to that of  the regular lattice, especially the case with $b=2$.

{As a general remark, from Table \ref{tab_ce_2d_tot} we observe that in
passing from $b=3$ to $b=5$, and thus increasing the connectivity of the lattice, 
for the pure model we obtain a slight  improvement for all exponents towards  the exact values of the square lattice, though the values are very far from the exact ones.
A possible convergence is, thus, so slow that 
  the degree of inner correlation of a folded cell of a HL necessary to reproduce 2D RG might be so large to make a single cell comparable with a whole  real Bravais lattices. }

{We now move to models in which there is a spin-glass phase, in order to test the critical behavior prediction of HL RG
in the case of strong disorder.}


\subsection{Random bond Ising model in $d\geq 3$}
\label{sec:I3D}
In this section we consider the Ising model with 
the $\pm J$ coupling distribution, Eq. (\ref{eq:PpmJ}) on hierarchical
lattices mimicking  the topology of the cubic lattice.
Recently, Salmon, Agostini and Nobre have studied the Ising spin glass  on the hierarchical Wheatstone bridge (WB) 
lattices \cite{Salmon10}, 
obtaining accurate phase diagrams and showing that on these lattices the lower critical dimension for the spin glass phase is greater than $ d=\ln 5/\ln 2\simeq 2.32 $, cf. the WB HL in Fig. \ref{diamond}.
The next pattern of the  WB family, cf. Fig. \ref{WB_lattice}, corresponding to a lattice in dimension higher than $2.5$ 
has a fractal dimension $ d \approx 3.58$.
  Consistent quantitive deviations from  the cubic lattice  behavior might, then,  occur.

\begin{figure}
\begin{center}
\includegraphics[width=\columnwidth]{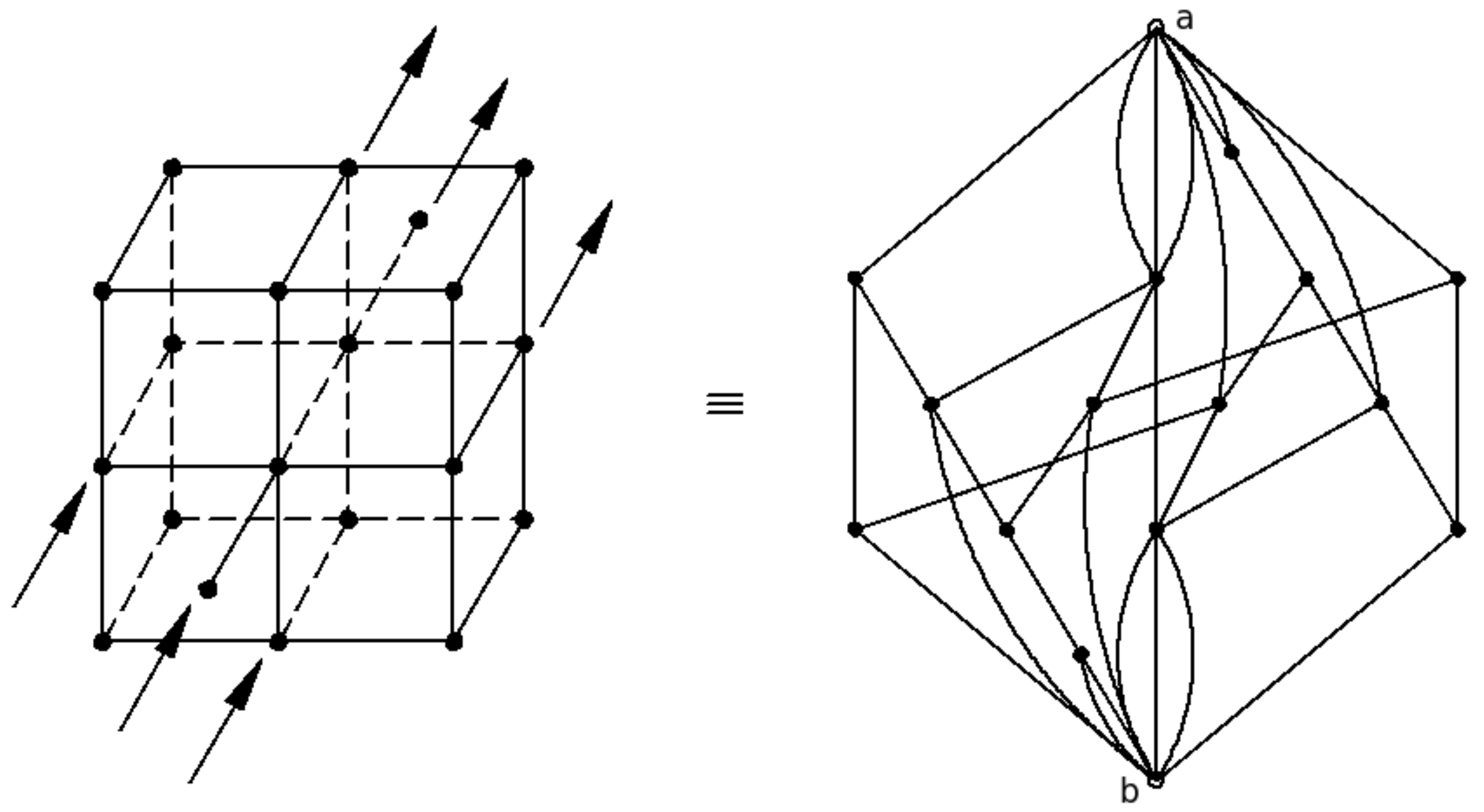}
\caption{Cubic cells of length $b=3$ on the left
  hand side, and relative folded cube hierarchical lattice on the right hand side.  The two
  roots are open circles.  All {\em incoming} sites and all
  {\em outcoming} ones, pointed by the arrows in the left figure, are put together and
  generate root sites {\tt{a}} and {\tt{b} }on the right hand side figure.  
  The inner sites of the cubic cell become the inner sites of the
  hierarchical cell.}
\label{Tsallis_3d}
\end{center}
\end{figure}

\begin{figure}
\begin{center}
\includegraphics[width=.55\columnwidth]{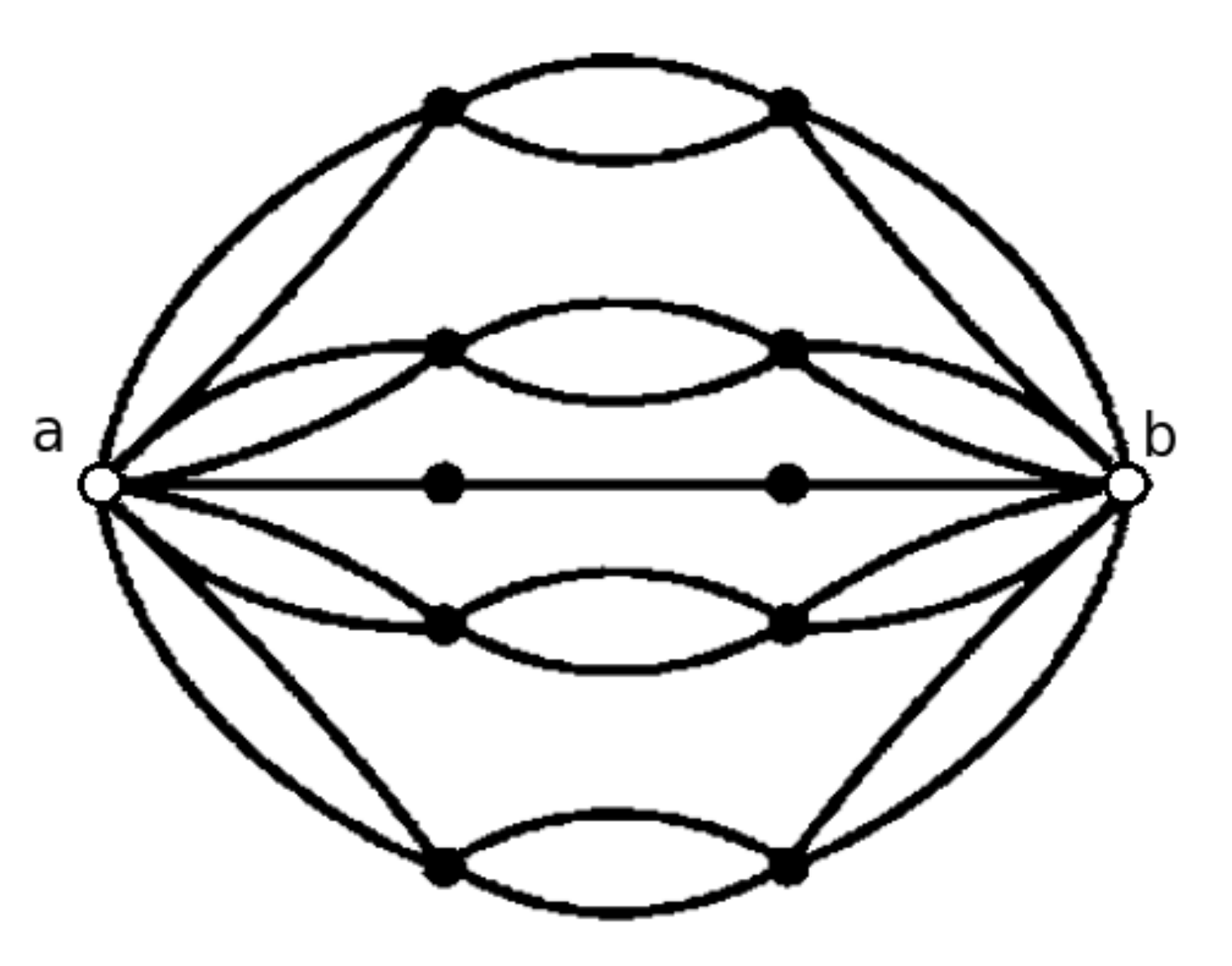}
\caption{MK lattice with $b=d=3$.\cite{Erbas05,Hinczewski05} }
\label{reticolo_MK_3d}
\end{center}
\end{figure}

We, here, investigate the model on the Folded Cube (FC) hierarchical lattice shown in Fig. \ref{Tsallis_3d}. 
This is the "three-dimensional" extension of the folded square lattice of Fig. \ref{Tsallis_2d_b3}, 
and it was introduced to study the anisotropic ferromagnetic Potts model  
in three dimensions \cite{Tsallis84}. This lattice has a fractal dimension $d = \ln 35/\ln 3 \simeq 3.2362$, 
always larger than $3$, though nearer to it as compared to the WB. It has $b = 3$ and, unlike the latter, it is able to retain a possible antiferromagnetic order, as well.
As a further comparison we will report on the critical properties of the Ising spin-glass model on  a MK
lattice with $b=3$  and fractal dimension $d=3$, cf. Fig.  \ref{reticolo_MK_3d},
introduced in Refs. \cite{Erbas05,Ozcelik08}.

\begin{figure}[t!]
\begin{center}
\includegraphics[width=\columnwidth, angle=0]{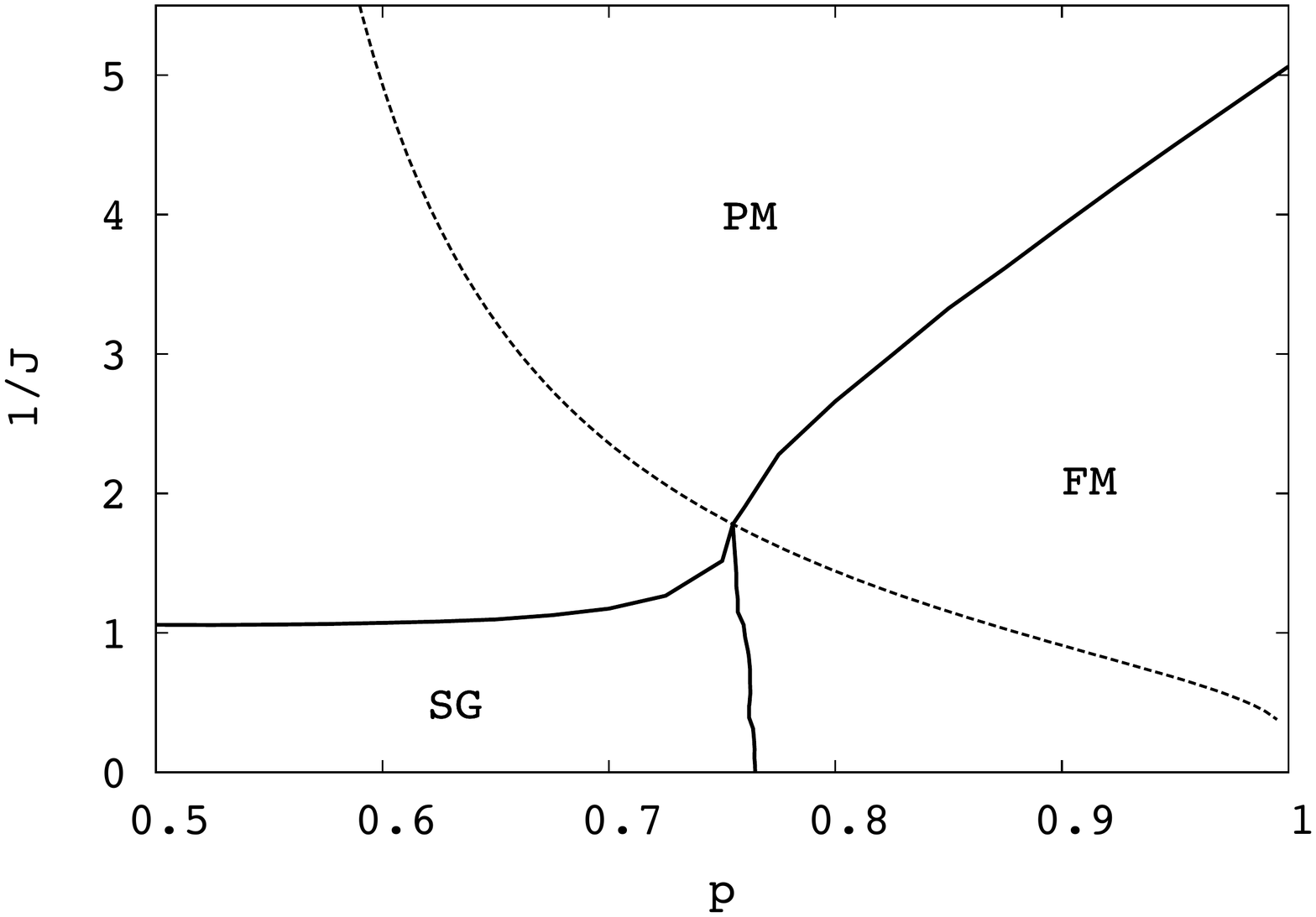}
\caption{Phase diagram of Ising spin glass model on the Folded Cube HL, cf. Fig. \ref{Tsallis_3d}, obtained for $M=5 \cdot 10^4$ and $N_S=10$. The  dashed line represents the Nishimori line. The plot is symmetric in $p\to 1-p$ and an antiferromagnetic phase is present a small $p$, in place of the FM one. }
\label{P-D_Tsallis_50k}
\end{center}
\end{figure}

The resulting phase diagram is shown in Fig. \ref{P-D_Tsallis_50k}.
An important feature of the phase diagram 
is the small reentrance in the region below the multicritical point.
As a consequence, by lowering the temperature
one can go from the high temperature (disordered) paramagnetic (PM) phase 
to an ordered ferromagnetic (FM) phase and, 
eventually to a low temperature disordered spin-glass (SG) phase. 
The values of the critical points are reported in Table ~\ref{tab_3d_CP}.

\begin{table}[t!]
\begin{center}
\begin{tabular}{c c c c c c}
\hline
 \phantom{.}&   \phantom{.}\qquad       $\begin{array}{c}p \\T\end{array}$ &\quad    fixed   &       points        &     coordinates    \qquad \phantom{.}      
\end{tabular}
\begin{tabular}{|c|c|c|c|c|}
\hline
 HL     & FM & SG  & $T=0$  & MC   \\
\hline
  $\begin{array}{c}\mbox{Fold cube}\\ \mbox{ Fig. \ref{Tsallis_3d} }\end{array}$
   				     &  $\begin{array}{c} 1 \\  5.066(1)\end{array}$
 				     &  $\begin{array}{c} 0.5 \\ 1.072(1) \end{array}$
 				     &  $\begin{array}{c} 0.764(2)  \\ 0   \end{array}$ 
				     &  $\begin{array}{c} 0.7547(3) \\ 1.779(1) \end{array}$     \\
                           \hline
$\begin{array}{c}\mbox{ WB ``3D"}\\ \mbox{ Fig. \ref{WB_lattice} }\end{array}$
 					 & $\begin{array}{c} 1 \\  5.457(1)\end{array}$
					 & $\begin{array}{c} 0.5 \\ 1.112(2) \end{array}$ 
					 & $\begin{array}{c} 0.760(1)  \\ 0   \end{array}$ 
					 & $\begin{array}{c} 0.745(2) \\ 1.620(2) \end{array}$\\
\hline
$\begin{array}{c}\mbox{ WB ``4D"}\\ \mbox{ Ref.  \cite{Salmon10} }\end{array}$
					  &  
					  & $\begin{array}{c} 0.5 \\ 2.515(2) \end{array}$
					  & $\begin{array}{c} 0.667(2)  \\ 0   \end{array}$  
					  & $\begin{array}{c} 0.664(2) \\ 2.836(2) \end{array}$  \\
\hline
$\begin{array}{c}\mbox{ MK}\\ \mbox{Fig. \ref{reticolo_MK_3d} }\end{array}$
 					& $\begin{array}{c} 1 \\ 5.383(1)  \end{array}$
					& $\begin{array}{c} 0.5 \\ 1.136(4) \end{array}$ 
					& $\begin{array}{c} 0.761(1)  \\ 0   \end{array}$  
					& $\begin{array}{c} 0.752(7) \\ 1.797(3) \end{array}$ \\
					\hline
\end{tabular}
\caption{Position of typical critical points of phase diagram in Fig. \ref{P-D_Tsallis_50k} for the hierarchical lattice \ref{Tsallis_3d}: ferromagnetic (FM),
spin-glass (SG), zero temperature ($T=0$) and multi critical point (MC). For comparison also values of the same fixed points are reported for other hierarchical lattices discussed in the text. }
\label{tab_3d_CP}
\end{center}
\end{table}

\subsubsection{FM fixed point}
The transition at $p = 1$ on  the folded cube cell is obtained at $ T_c = 5.066(1) $, $12\%$ larger as compared 
to the same model on cubic lattice (where $ T_c = 4.5115... $ \cite{Talapov96}).
On the $d\simeq 3.58$ WB lattice the difference was about $21 \%$. \cite{Salmon10}
On the MK lattices the best known result is obtained for the $d=3$ lattice in Fig. \ref{reticolo_MK_3d}, where the critical temperature 
of the pure transition is $ T_c = 5.38(3) $, $19\%$ larger than the one in the cubic lattice. \cite{Erbas05}

In the folded cube the points on the critical line between the FM  and the
PM phases are attracted to a single fix point located at $p = 1$. 
The critical exponents at this point are $ y_T =  1.523(1)$ and $ y_h = 1.864(1)$, 
identifying a second order transition \cite{Nienhuis77}.
On other HLs we obtain: $y_T=1.149(1)$ and $y_h=0.9636(1)$ for the WB lattice in Fig.~\ref{WB_lattice}; 
$y_T=1.460(1)$ and $y_h=1.613(1)$ for the MK lattice in Fig.~\ref{reticolo_MK_3d}.
These are to be compared with numerical estimates  by means of  simulations on the cubic lattice:  
$y_T =1.587(1) $ and $ y_h=2.482(1) $  \cite{Pelissetto02}. 
None of them agrees with the cubic lattice results, even  though the FC lattice in Fig.~\ref{Tsallis_3d} 
yields the nearest estimate for both exponents.
The corresponding physical exponents are reported in Table \ref{tab_ce_3d_pure}.
Note, in particular, that for all the HLs considered here we have $ \alpha <0 $. {According to the Harris criterion \cite{Harris74}
this  would imply that disorder is irrelevant in modifying the ferromagnetic critical behavior, whereas for all these HL's one finds a
frozen phase different from the ferromagnetic one in a given interval of $p$ values at low temperature: besides the "weak disorder" 
ferromagnetic fixed point a second ``strong disorder" spin-glass fixed point arises in presence of quenched randomness.}

\begin{table}[t!]
\begin{center}
\begin{tabular}{|c|c|c|c|c|}
\hline
 critical& MK & WB & Folded cube & Cubic \\
 index & {\footnotesize {Fig. \ref{reticolo_MK_3d} }} & {\footnotesize{Fig. \ref{WB_lattice}}} & {\footnotesize{Fig. \ref{Tsallis_3d}}} & {\footnotesize {Ref. \cite{Pelissetto02}}} \\
\hline
 $\alpha$  & $-0.2169(1)$   & $-1.121(1)$ & $-0.1253(1)$ & $0.110(1)$\\
 $\beta $  & $1.112(1)$     & $2.282(1)$   & $0.9012(1)$  & $0.3265(3)$ \\
 $\gamma$  & $-0.006490(1)$ & $-1.443(1)$  & $0.3229(1)$  & $1.2372(5)$ \\
 $\delta$  & $0.9942(1)$    & $0.3676(1)$  & $1.358(1)$  & $4.789(2)$ \\
 $\nu   $  & $0.6850(1)$    & $0.8705(1)$  & $0.6567(1)$  & $0.6301(4)$ \\
 $\eta $   & $2.009(1)$     & $3.658(1)$   & $1.508(1)$  & $0.0364(5)$ \\ \hline
\end{tabular}
\caption{Critical exponents in pure ferromagnetic critical point for Ising model on the folded cubic cell and comparison.  } 
\label{tab_ce_3d_pure}
\end{center}
\end{table}


\subsubsection{Spin-glass fixed point}

For the FC HL in the totally disordered case $ p = 0.5 $ the PM-SG transition is located at $ T_c = 1.072(1)$, 
with a difference of $4.5\%$ relatively to the Bravais cubic lattice critical point, 
for which the Monte Carlo simulations provide $ T_c = 1.120 (4) $ \cite{Montecarlo_3d}.
As reported in Table  \ref{tab_3d_CP} the WB lattice in Fig. \ref{WB_lattice}
has $ T_c = 1.112 (2) $ and the MK lattice in Fig. \ref{reticolo_MK_3d} has $ T_c  = 1.136(4)$.

\begin{figure}[b!]
\begin{center}
\includegraphics[width=\columnwidth]{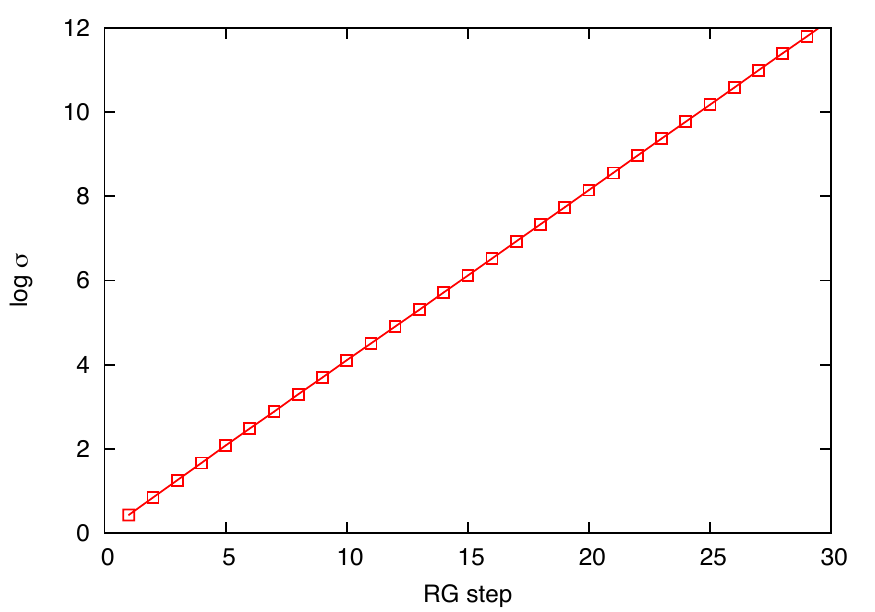}
\caption{Graph of $ \ln (\sigma_J) $ under repeated applications of the PSRG transformation
 at zero temperature and $p=0.5$ on lattice in Fig. \ref{Tsallis_3d}.}
\label{logJ_vs_n}
\end{center}
\end{figure}

The behavior of $ \ln (\sigma_J) $ under successive PSRG iterations for $ p = 0.5 $ is shown in Fig. \ref{logJ_vs_n}. 
From its linear behavior we estimate for the stiffness exponent of the spin-glass phase 
 $ \theta =  0.2052(1)$ (cf. Sec.~\ref{Sec:Num_res_RB_2d}). 
For the cubic lattice one finds in the literature $ \theta = 0.19(1)$ \cite{Bray87}  or $ \theta = 0.20(5)$ \cite{Katzgraber2008}.
We, thus, obtain a result very close to that expected for the cubic lattice.  
On the WB lattice one has $ \theta = 0.297(3)$ \cite{Salmon10}, while for the MK lattice  it is 
found $ \theta \approx 0.27 $. \cite{Katzgraber2008} The result are summarized in Table \ref{tab_stiffness_3d}.

\begin{table}
\begin{center}
\begin{tabular}{|c|c c|c|}
\hline
       Lattice   type & Fig.&Ref.   & $\theta$ \\
       \hline
MK $d=3$ & \ref{reticolo_MK_3d} & \cite{Katzgraber2008}& $0.27(1)$\\
WB $d\sim 3.58$       &\ref{WB_lattice}  & \cite{Salmon10}  &$0.297(3)$\\
FC $d\sim 3.24$       &  \ref{Tsallis_3d} &&$0.2052(1)$\\
\hline
Cubic &&\cite{Bray87} & $0.19(1)$\\
Cubic & &\cite{Katzgraber2008} & $0.20(5)$ \\
\hline
\end{tabular}
\caption{Stiffness exponent as computed on different HL and different estimates on the cubic lattice.}
\label{tab_stiffness_3d}
\end{center}
\end{table}

\begin{figure}[t!]
\begin{center}
\includegraphics[width=\columnwidth]{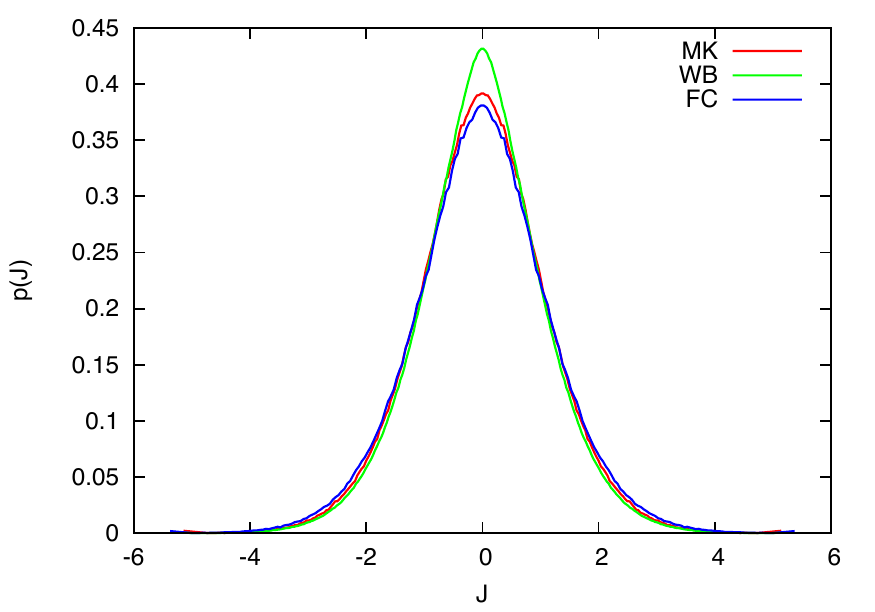}
\caption{Fixed probability distribution for the spin glass - paramagnet transition for Ising spin glass on lattice in Fig. \ref{reticolo_MK_3d} (MK), 
Fig. \ref{WB_lattice} (WB) and Fig. \ref{Tsallis_3d} (FC). 
These distributions are reached after about $2-3$ steps of renormalization,
starting from the binomial distribution, and maintain this shape for about $10-15$ steps steps, after which they starting moving away because of  statistical fluctuations. }
\label{all_FD}
\end{center}
\end{figure}

All critical points along the PM-SG transition are attracted by an unstable fixed point at $p=0.5$.
In Fig. \ref{all_FD} we show the unstable fix point coupling distributions for the three 
lattices considered so far. 
Critical exponents for this fix distribution are computed by adapting the methods used for the RFIM in Sec. \ref{sec:RFIM}.
A first exponent $y_h$ can be obtained generalizing the pure case definition: $\lambda_h=\langle \partial_h h_R \rangle$, 
where the average is carried out by extracting the interactions from the fix point distribution. 
Indeed, in our case this is easy and can be exactly calculated. In fact for the fix
point distribution we have $h=h^\dagger=0$, and so $\mathcal{H}(s)=\mathcal{H}(-s)$. 
It is, then, straightforward to obtain, using Eqs. (\ref{eq_JdJ}) and (\ref{eq_der_BF}) 
that $\partial_h h_R= c$ for every choice of the $J$ interactions, where $c$ is the number 
of internal sites connected to each external site: $c=9$ for the folded cube cell in Fig. \ref{Tsallis_3d} 
and the MK in Fig. \ref{reticolo_MK_3d}, $c=4$ for the WB in Fig. \ref{WB_lattice}.  
For all these cells the exponent $y_h$ is then $y_h=\log_b \langle \partial_h h_R \rangle = 2$ (see Sec. \ref{ss:FMIM}).
We note that $y_h < d$, ensuring that the magnetization is continuous at the transition (no first order transition).

To get the correlation length exponent $\nu$ we note that
defining $t \equiv \sigma_J - \sigma_J^*$, where $\sigma_J^*$ is the value at the critical point, 
we obtain the scaling law
\footnote{note that, as always in this paper, we are using the reduced variables $ \beta J \rightarrow J$}
\begin{equation}
 \xi (t) = b~  \xi (t') = b ~ \xi (b^x t) = b^n  \xi (b^{nx} t).
\end{equation}
By taking $n$ such that $b^{nx} t = t_0$, where $t_0$ is arbitrary but fixed, we end up with
\begin{equation}
 \xi (t) = \biggl( \frac{t_0}{t} \biggl) ^{\frac{1}{x}} \xi (t_0) \sim t^{-\frac{1}{x}} ,
\end{equation}
from which we get $\nu = 1/x$:
the value of $\nu$ can be estimated by the trend of $\sigma_J$ near the critical point distribution.

We use  the method already exposed in Sec. \ref{sec:RFIM}: after obtaining an instance of the fixed distribution, 
we make a copy of it and multiply  each coupling of the copy  by a quantity $ 1 + \delta$, with  $\delta \approx 0.05 $.
The two copies of the distribution are, then, simultaneously iterated in the PSRG transformation, 
with the first copy forced near the fixed point and the second one free to flow.
We eventually estimate $\nu$ by means of Eq. (\ref{nu_tn}),
where the parameter $ t_n $ is now the difference between the  values of  $ \sigma_J $ in the two copies at RG step $n$.

Continuing  this way we find the values shown in Table \ref{table_nu} for different HLs and we notice that they  are
compatible with each other within the statistical error. 
However, when compared to the estimate for the cubic Bravais lattice, none of them is compatible with it:
$ \nu = 2.45(15) \rightarrow 1 / \nu = 0.408(25) $ \cite{Pelissetto08}.
 On top of that, we observe that quantitatively, the FC cell, of fractal dimension $3.24$ has a value of $\nu$ further away than 
 the value on the MK and WB ones. This is a strong signature of the limitations  of the RG approach on  HL's to the critical behavior of systems on Bravais lattices in presence of disorder.



Proceeding as  for the correlation length, and using the free energy $f$ scaling law
\begin{eqnarray}
 f (t) = b^{-d}  f(t') = b^{-d}  f (b^x t) = b^{-nd}  f (b^{nx} t)
 \nonumber
\end{eqnarray}
with $n$ such that $b^{nx} t = t_{0}$ and $t_{0}$ arbitrary but fixed, we have
\begin{equation}
 f (t) = \biggl( \frac{t_{J}}{t_{0}} \biggl) ^{\frac{d}{x}}  f (t_{0}) \sim t^{\frac{d}{x}}.
\end{equation}
Then from $\partial^2 f/\partial t^2 \sim |t|^{-\alpha} $, we obtain the scaling relation
\begin{equation}
 \alpha = 2 - \frac{d}{x} = 2-d\nu,
\end{equation}
as well as
\begin{eqnarray}
 \beta &=& \frac{\nu(1+\eta)}{2} \, ,\nonumber \\
 \gamma &=& (2-\eta)\nu \, , \nonumber
\end{eqnarray}
where 
\begin{equation}
 \eta = d+2-2 y_h = d-2 \,  \nonumber 
\end{equation}
since $y_h=2$ for all our HLs.
The estimates of the physical exponents are reported in Table \ref{table_pe_SGP}.

\begin{table}[t!]
\begin{center}
\begin{tabular}{|l|c|c|c|c|}
\hline
Lattice & Fig. & Resc.& Dim. & $1/\nu$ \\
\hline
MK & \ref{reticolo_MK_3d}    &$ b=3$ &$ d=3$ & $0.297 \pm 0.026$ \\
WB &  \ref{WB_lattice}        &$ b=2$ & $d\sim3.58$ & $0.308 \pm 0.063$ \\
Folded cube &  \ref{Tsallis_3d}   &$ b=3$ & $d\sim3.24$ & $0.262 \pm 0.037$\\
Bravais cubic \cite{Pelissetto08} & & & $d=3$ & $0.408 \pm 0.025$ \\ \hline
\end{tabular}
\caption{Estimates for the exponent $ \nu $ at transition SG-para for the Ising spin glass model.}
\label{table_nu}
\end{center}
\end{table}

\begin{table}
\begin{center}
\begin{tabular}{|c|c|c|c|c|}
\hline
 critical& MK & WB  & Folded cube & Cubic \\
 index & {\footnotesize {Fig. \ref{reticolo_MK_3d} }}& {\footnotesize{Fig. \ref{WB_lattice}}} &  {\footnotesize{Fig. \ref{Tsallis_3d}}} & {\footnotesize {Ref. \cite{Pelissetto08}}}\\
\hline
 $\alpha$  & $-8.10(88)$ & $-9.6(24)$ & $-10.4(17)$ & $-5.4(5)$ \\
 $\beta $  & $3.37(30)$  & $4.20(86)$  & $4.27(60)$  & $0.77(5)$ \\
 $\gamma$  & $3.37(30)$  & $1.35(28)$  & $2.92(41)$  & $5.8(4)$ \\
 $\nu   $  & $3.37(30)$  & $3.25(66)$  & $3.82(54)$  & $2.45(15)$ \\
 $\eta $   & $1$  & $1.585\ldots$  & $1.236\ldots$  & $-0.375(10)$ \\ \hline
\end{tabular}
\caption{Estimates for the physical exponents at the SG-PM transition for the Ising spin glass model.  The 
exponent $\eta$ is $d-2$, $d$ being the fractal dimension of the HL.}
\label{table_pe_SGP}
\end{center}
\end{table}

Concluding this section, on the FC lattice in Fig. \ref{Tsallis_3d} the estimates for the 
pure criticality are much closer to those on cubic lattice (although still not compatible), 
compared to MK in Fig. \ref{reticolo_MK_3d} and WB in Fig. \ref{WB_lattice}.
Also for the stiffness exponent of SG phase a remarkable improvement is observed: 
its estimate is compatible with its cubic value  on the FC lattice
but not on  other HLs. 
For other quantities, though, at the disordered fixed point such improvement
is unseen. In particular, in the estimate for the $\nu$ exponent at the SG-PM transition.

We can argue that  the stiffness exponent depends more on the local geometrical properties of the lattice, that are 
substantially improves in the folded cube lattice, while critical properties depend
 on longer distances, that are still dominated by the hierarchical backbone an thus far from the Bravais lattice behavior.
 
More stringent tests can be obtained for the Blume-Emery-Griffiths model, that we will analyze in the next section. 
For this system the inverse first order transition expected from mean-field theory \cite{Crisanti05b} and simulation in finite dimension \cite{Paoluzzi10} 
is absent on the MK lattice in Fig. \ref{reticolo_MK_3d}. \cite{Ozcelik08}


\section{Blume-Emery-Griffiths model}
\label{sec:BEG3D}
We now move on to a different system, the Blume-Emery-Griffiths (BEG) model,
originally devised to study the superfluidity transition and phase
separation in He$^3$-He$^4$ mixtures \cite{Blume71}. The model is
known to display, besides a second order phase transition, also a
first order transition, both in the ordered case 
(between the PM and FM phases) and in the case with 
quenched disordered interactions (between
the PM and SG phases). 

The ordered model cases have been
introduced and solved in the mean-field approximation in
Refs. \cite{Blume66,Capel66,Blume71}.  Finite dimensional analysis has
been carried out by different means, e.g., series extrapolation
techniques \cite{Saul74}, PSRG analysis \cite{Berker76}, Monte Carlo
simulations \cite{Deserno97}, effective-field theory
\cite{Chakraborty84} or two-particle cluster approximation
\cite{Baran02}.

Extensions to quenched disordered models, both perturbing the
ordered fixed point and in the regime of strong disorder, have been studied
throughout the years by means of mean-field theory \cite{Crisanti02,Crisanti04, Crisanti05},
 PSRG analysis on
Migdal-Kadanoff hierarchical lattices \cite{Falicov_96,Ozcelik08}, and 
Monte Carlo numerical simulations \cite{Puha00,Paoluzzi10,Paoluzzi11,Leuzzi11}.
The latter studies show that a critical transition line separates the SG and 
PM phases. 
Like in the mean-field cases, it consists of a second order transition terminating in  
a tricritical point  from which a first order inverse transition starts. \cite{Paoluzzi10}
Furthermore, a reentrance of the first order transition line 
is present for positive, finite values of the chemical potential of the holes, \cite{Crisanti05} 
yielding the so called inverse freezing phenomenon of an amorphous phase  arresting itself in a blocked, solid-like state
upon heating.

\begin{figure}[t!]
\begin{center}
\includegraphics[width=.45\textwidth]{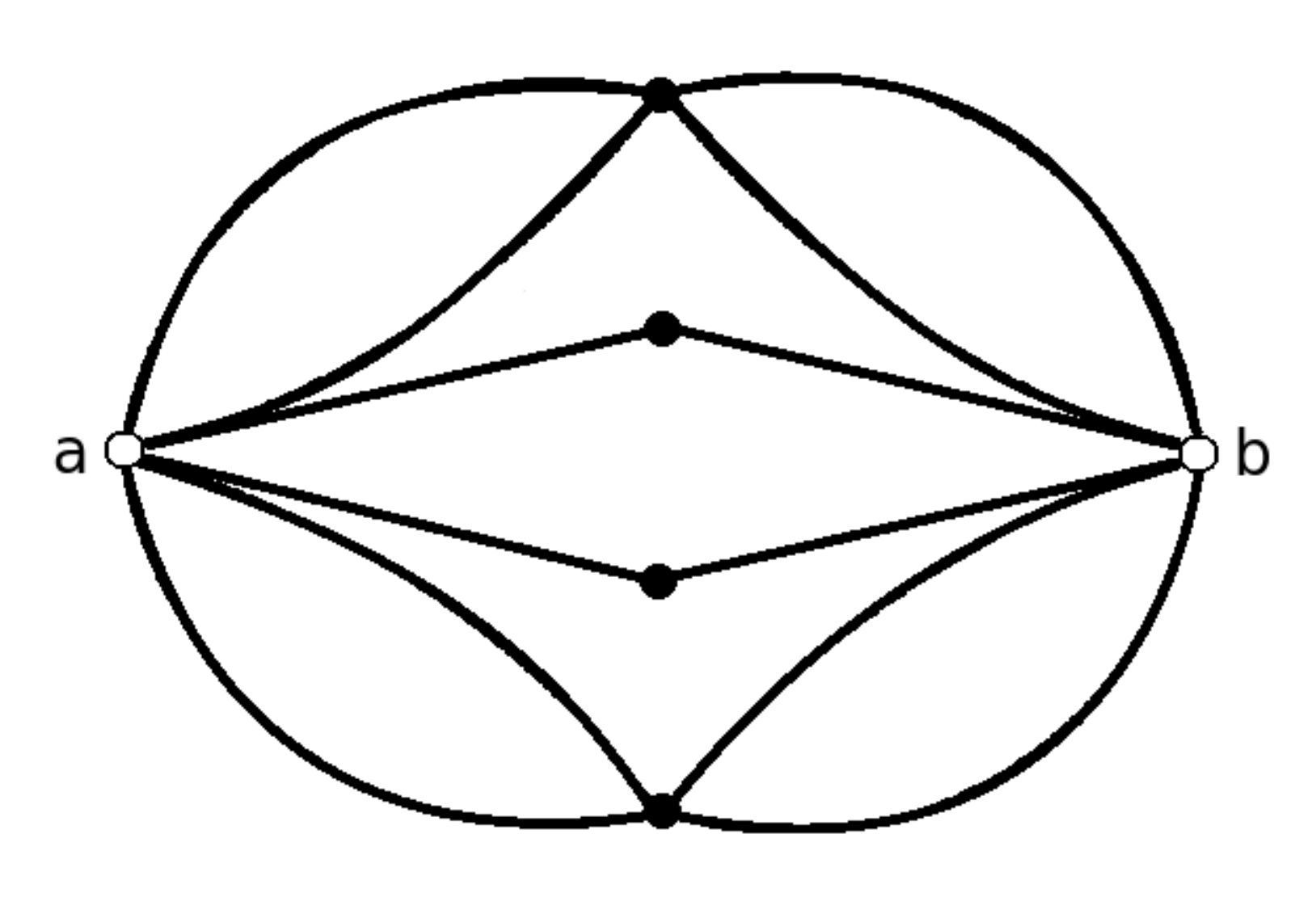}
\caption{MK lattice with fractal dimension $d=3.58$, the same as WB lattice in Fig. \ref{WB_lattice}. }
\label{lattice_MK_358}
\end{center}
\end{figure}

In absence of any purely ferromagnetic contribution, 
the PSRG approach on MK cells apparently does not show any first order phase transitions, nor any reentrance, 
as shown by Ozcelik and Berker \cite{Ozcelik08}. 
In this section we will deepen their analysis at $p=1/2$, investigating 
the critical behavior on the MK lattices of Fig. \ref{reticolo_MK_3d} and Fig. \ref{lattice_MK_358}, 
on the Wheatstone Bridge (WB) lattice of Fig. \ref{WB_lattice} and on the Folded Cube (FC) lattice of Fig. \ref{Tsallis_3d}.

The BEG model with generic magnetic exchange interaction is defined by the Hamiltonian
(we use  reduced variables)

\begin{equation}
-  \H(s) =  \sum_{ <ij> } J_{ij}s_i s_j + K \sum_{ <ij> } s^2_i s^2_j - \Delta \sum_i s^2_i \, ,
\end{equation}
where $s_i = \pm 1, \, 0$ and $J_{ij}$ can be deterministic or quenched disordered and distributed according to some probability distribution.
In the latter case, under PSRG transformation all renormalized interactions become quenched random and it is convenient to start the iteration using the more general form
\begin{eqnarray}
&& -  \mathcal{H}= \sum_{\left\langle ij \right\rangle}  J_{ij}s_i s_j 
+ \sum_{\left\langle ij \right\rangle} K_{ij} s^2_i s^2_j 
\nonumber 
- \sum_{\left\langle ij \right\rangle} \Delta_{ij} \left( s^2_i + s^2_j \right) 
- \sum_{\left\langle ij \right\rangle} \Delta_{ij}^{\dagger} \left( s^2_i - s^2_j \right)
\\ && 
\end{eqnarray}
The model is, further, defined, by the multivariable initial probability distribution of the interactions:
\vspace{0.2cm}
\begin{align*}
 {\cal P}(J_{ij},K_{ij},\Delta_{ij},\Delta^{\dagger}_{ij}) \!= && \!\!\!\! \frac{\delta(J_{ij}-J_0)+\delta(J_{ij}+J_0)}{2} 
\delta(K_{ij}) ~ \delta(\Delta_{ij}-\Delta_0) ~\delta(\Delta_{ij}^{\dagger}) \, . \\
\end{align*} 
We notice that if $\Delta_0 \ll -1$, the values $s_i=0$ are
suppressed and the model tends to the Ising spin glass model analyzed in previous sections:
\BEA
&&-  \H(\{J_{ij}\};\{s_i\})= \sum_{\langle ij\rangle}J_{ij}s_is_j
\nonumber 
\EEA
with $s_i = \pm 1$ and
\begin{equation}
 P(J_{ij}) = \int dK_{ij} \, d\Delta_{ij} \, d \Delta^\dagger_{ij} \,{\cal P}(J_{ij},K_{ij},\Delta_{ij},\Delta^{\dagger}_{ij}) \nonumber \, .
\end{equation}

Decimating the inner sites at a given hierarchical cell with fixed outer sites $s_a$ and $s_b$, 
and using the up-down symmetry of the Hamiltonian,
the relations for the renormalized interactions imposed by the conservation of 
the partition function, cf. Eq. (\ref{eq:partition}), 
can be written similarly to Eqs. (\ref{eq:partition})-(\ref{eq_Ising}) as

\begin{eqnarray}
 J_R &=& \frac{1}{2} \log \biggl( \frac{x_{++}}{x_{+-}} \biggl) , \, \nonumber 
\\
 K_R &=& \frac{1}{2} \log \biggl( \frac{x_{++} ~ x_{+-}  ~ x_{00}^2}{x_{0+}^2  ~ x_{+0}^2} \biggl) , \, \nonumber
\\
 \Delta_R &=& \frac{1}{2} \log \biggl( \frac{x_{00}^2}{x_{+0} ~ x_{0+}} \biggl) \, , 
\\
 \Delta^{\dagger}_R &=& \frac{1}{2} \log \biggl( \frac{x_{+0}}{x_{0+}} \biggl) \, ,  \nonumber
\end{eqnarray}
where $x_{s_as_b}$ are the edge Boltzmann factors, cf. Eq. (\ref{eq:partition}).
In the $T=0$ limit these relations become 
\begin{eqnarray}
 2 J_R &=& \max\bigl[- \mathcal{H}(1,1,s)\bigr]
 -\max\bigl[- \mathcal{H}(1,-1,s)\bigr]
  \nonumber
\\
2 K_R &=& \max\bigl[- \mathcal{H}(1,1,s))\bigr]
+ \max\bigl[- \mathcal{H}(1,-1,s)\bigr]
 \nonumber 
 \\
 &&-2\bigr\{ \max\bigl[- \mathcal{H}(1,0,s)\bigr]- \max\bigl[- \mathcal{H}(0,1,s)\bigr] \bigr\}
  \nonumber
 \\
 &&+2\max\bigl[- \mathcal{H}(0,0,s)\bigr]
\label{eq_BEG_zeroT}
\\
 2\Delta_R &=&2 \max\bigl[- \mathcal{H}(0,0,s)\bigr]\nonumber 
 \\
 &&- \max\bigl[- \mathcal{H}(1,0,s)\bigr] 
 - \max\bigl[- \mathcal{H}(0,1,s)\bigr] \nonumber
\\
 2\Delta^{\dagger}_R &=&\max\bigl[- \mathcal{H}(1,0,s)\bigr] 
 - \max\bigl[- \mathcal{H}(0,1,s)\bigr]
  \nonumber
\end{eqnarray}

\begin{figure}[t!]
\begin{center}
\includegraphics[width=\columnwidth, angle=0]{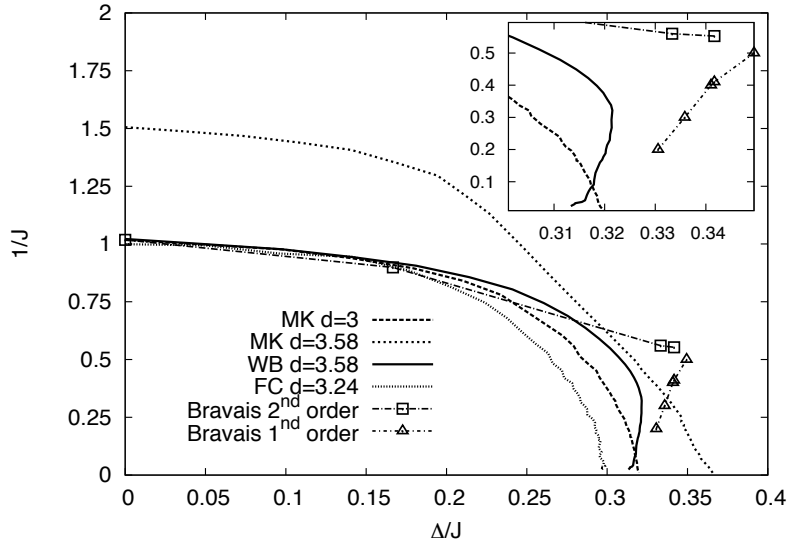}
\caption{Phase diagram for the BEG model on the MK$(d=3)$ lattice in Fig. \ref{reticolo_MK_3d}, WB lattice in Fig. \ref{WB_lattice},
FC lattice in Fig. \ref{Tsallis_3d} and simulations \cite{Paoluzzi10}. The line for the MKs and WB are obtained with $M=10^6$, whilst for the FC we use $M=5 \cdot 10^4$. 
Inset: detail of the reentrance region for WB  and 3D cubic lattices, compared to the MK HL line (no reentrance).
}
\label{fig_pd_all_BEG}
\end{center}
\end{figure}

The phase diagrams relative to the different hierarchical lattices are shown in Fig. \ref{fig_pd_all_BEG}.
They are obtained representing the probability distributions by a pool of $M=10^5$ interaction quadruples
 $(J,K,\Delta,\Delta^\dagger)$ and the RG evolution of each distribution is analyzed over $N_S=10$
different samples, i.e., starting with ten different initial realizations of the quenched couplings. 
As worked out in Sec. \ref{sec:Ising}, paramagnetic, ferromagnetic and spin glass phases 
are determined by the analysis of the PSRG flux of $J= \langle J_{ij} \rangle$ and $\sigma^2_J = \langle J_{ij}^2 \rangle-\langle J_{ij}\rangle^2$.

\begin{figure}[t!]
\hspace*{-.5cm}\includegraphics[width=1.15\columnwidth]{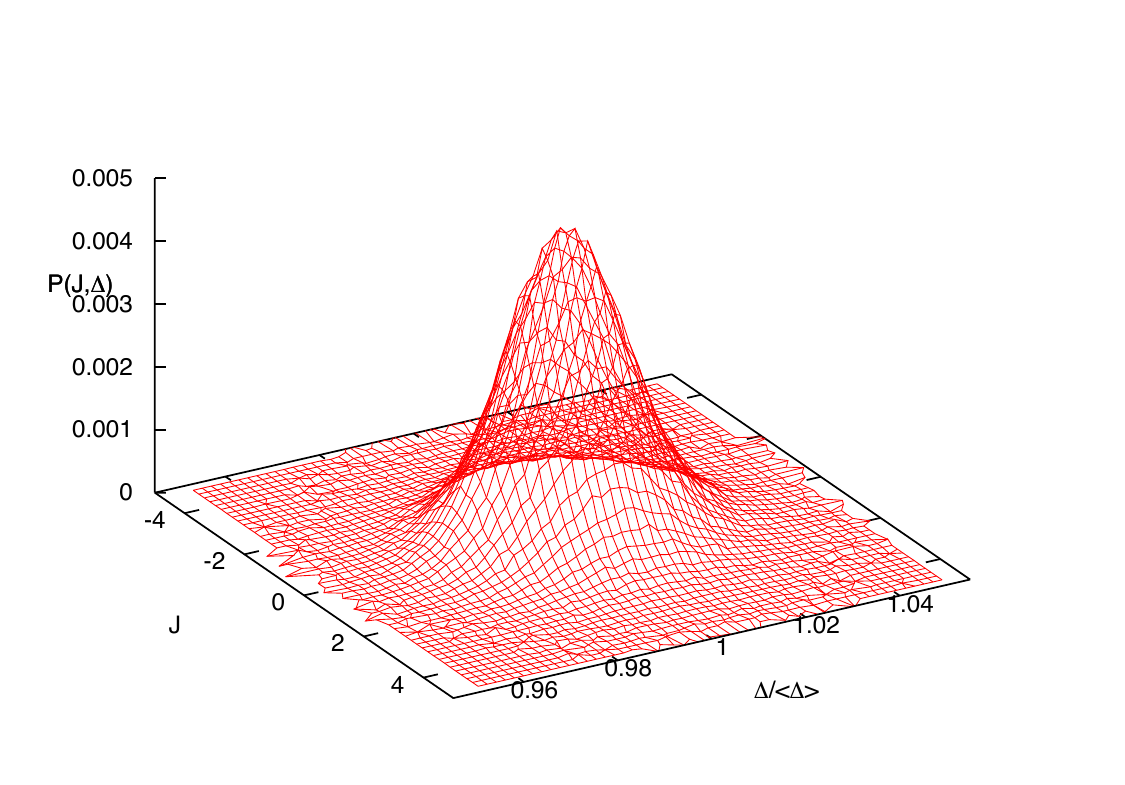}
\caption{(Projection of unstable fixed distribution ${\cal P}(J,K,\Delta,\Delta^\dagger)$ for BEG model at the transition SG-para on the WB lattice in Fig. \ref{WB_lattice}.
In the distribution $\Delta$ is runaway to minus infinity exponentially, while  other interactions remain finite and the distribution retains this shape.  
The shape of relative distribution on the MK and FC lattices is very similar. }
\label{fig_fd_BEG}
\end{figure}

In all  cases we obtain a second order transition between paramagnetic and spin glass phases, with all points
on the transition line attracted by a unique fixed distribution at $\Delta \rightarrow - \infty$ (see Fig. \ref{fig_fd_BEG}), 
thus belonging to the same universality class of the  SG-PM transition in the Ising spin glass investigated in the previous section.

\subsection{Lack of first order phase transition}
Studying hierarchical lattices also much more complex than MK, 
the first order transition typical of the BEG model on Bravais lattice is not found, 
so we have a strong indication that this is an intrinsic limit of  hierarchical lattices, 
and not only of the MK kind of cells.

We have no clear evidence on why the first order transition is missing on the hierarchical lattices lattices we have considered so far. Based on physical arguments we may propose the following hypothesis. 
Second order transition are associated with an instability, the high temperature phase becomes unstable and a new stable phase appears. 
This instability can manifest itself locally, and hence hierarchical lattices can show a second order transition. 
First order transitions, on the contrary, are not triggered by an instability. The high temperature phase remains stable, 
but a thermodynamically more favorable phase takes over. Such a situation requires some sort of long range structure of the lattice, 
that is missing in the present hierarchical lattices, and this may explain why we do not see first order transitions. 
If our conjecture is correct we may wonder if the first order transition could still appear in hierarchical lattices with folded cells for $b$ finite but large enough. 
However, from our present knowledge this value of $b$ might be so large to make the HL approach ineffective. 
The analysis of the arising of a possible fist order critical behavior in $b$ is left for future work.

\subsection{Inverse freezing}

An apart feature of the BEG model arises, instead, adopting WB cells: 
the inverse transition between spin glass and paramagnet.
it occurs in the WB $d \approx 3.58$ where evidence for the reentrance is clearly obtained using the $T=0$ Eqs. (\ref{eq_BEG_zeroT}).
Inverse freezing is predicted in mean-field theory\cite{Crisanti05} and found in 3D numerical simulations.\cite{Paoluzzi10}.
It is not found in  MK lattices, instead,
neither in fractal dimension $3$ (cf. Fig. \ref{reticolo_MK_3d}) 
nor in $d\approx 3.58$ (cf. Fig. \ref{lattice_MK_358}) nor in the FC lattice.

\section{Conclusions}

We have provided a critical review of the standard methods to develop PSRG on hierachical lattices and applied them 
to different models.
On one side this analysis can be useful to test general results (e.g. Nishimori conjecture, Harris criterion),
and on the other side it is far easier and faster than Monte Carlo simulations on Bravais lattice. 
We, however, stress that these methods
display several drawbacks that we critically underlined model by model.

We have investigated the Random Field Ising Model (RFIM), the Random Bond Ising Model (RBIM) 
and the Blume-Emery-Griffiths (BEG) model defined on several MK and non-MK hierachical lattices, 
obtaining phase diagrams and critical exponents.

The RFIM has been analyzed on a non-MK (WB in Fig. \ref{WB_lattice})
and we show that the bimodal and Gaussian disordered cases belong to the same universality class.

The RBIM in $d<2.5$ has been analyzed on the folded square HLs family for $b=3$ (Fig. \ref{Tsallis_2d_b3}) and $b=5$ (Fig. \ref{Tsallis_2d_b5}),
where we find that the Nishimori conjecture fails and the disorder is irrelevant.

The RBIM in $d>2.5$ has been analyzed on  the folded cube (Fig. \ref{Tsallis_3d}) and compared to MK (Fig. \ref{reticolo_MK_3d}) 
and WB (Fig. \ref{WB_lattice}), where we obtain the critical exponents for the SG-PM transition and also show that the Harris criterion  turns
out to be satisfied.

The BEG model has been analyzed on non-MK HLs, i.e., WB (Fig. \ref{WB_lattice}) and folded cube (Fig. \ref{Tsallis_3d}), and 
compared to MK with $d=3$ (Fig. \ref{reticolo_MK_3d}) and $d\sim 3.58$ (Fig. \ref{lattice_MK_358}), and in all the cases the first order transition
taking place in the Bravais lattice for large enough chemical potential \cite{Crisanti05,Paoluzzi10} is absent. 

Our results provide a clue to the possibility of obtaining approximations of models on regular lattice
by  similar models on hierachical lattice.
We show that it is possible to obtain a good picture of the actual phase diagram, 
but far more difficult to yield a proper determination of critical exponents.

By introducing more complex elementary cells, with some non trivial internal structure, one hopes of 
capturing the local geometrical properties of the bonds. At least part of it. 
For pure models, although it is not a systematic approximation, 
a general improvement is obtained using unit cells that locally
mimic better the connectivity of the  Bravais lattice.
In the disordered case, instead, the situation is less definite, and no net improvement is observed:
the WB cell (Fig. \ref{WB_lattice}) proves to be the slightly most reliable 
(generally quantitatively better than the more complex folded cube in Fig. \ref{Tsallis_3d}), and in particular shows the
expected inverse transition for the BEG model, but we cannot give a general explanation for this.

In particular, the fractal dimension seems to play a minor role,
as the three dimensional regular lattice 
is better approximated by the WB with $d\sim 3.58$,
compared to $d=3$ MK (Fig. \ref{reticolo_MK_3d}) and $d\sim3.24$ folded cube (Fig. \ref{Tsallis_3d}),
while the $d\sim 3.58$ MK in Fig. \ref{lattice_MK_358} is the worst approximation by far.

The scaling factor $b$ has the known role to determine if the antiferromagnetic order 
can be preserved, as it is possible only when $b$ is odd so that negative interactions in the unit cell
involve negative interactions between the external sites 
(and this leads to a phase diagram symmetric under the inversion of the bonds sign).
Our results indicates that this feature does not play a crucial role in the disordered systems 
(at least until the negative bonds become dominant). The two investigated HLs with an even $b$, the WB and the $d\sim 3.58$ MK, indeed, appear to be, respectively, 
the best and the worst in approximating the phase diagram of the  models on regular lattice.
The only case in which the folded cube lattice in Fig. \ref{Tsallis_3d} provides a remarkable improvement with respect to the WB lattice
is in the estimate of the SG stiffness exponent. A more structured inner local connectivity
could, thus,  become important at low temperatures.

In conclusion, our results show that the approximation on HL is particularly poor for disordered systems, 
and strongly suggest that its limitations are intrinsic to the hierarchical nature. 
The most striking case is the lack of first order transition in BEG.
Indeed, using more complex HLs we may have a better treatment of short distances, i.e., short loops,  
but longer distances appear definitely dominated by the hierarchical backbone.

\begin{acknowledgements}

The research leading to these results has received funding
from  the People Programme (Marie Curie Actions) of the European Union's Seventh Framework Programme FP7/2007-2013/ 
under REA grant agreement n¡ 290038, NETADIS project and 
from the Italian MIUR under the Basic
Research Investigation Fund FIRB2008 program, grant
No. RBFR08M3P4, and under the PRIN2010 program, grant code 2010HXAW77-008.

\end{acknowledgements}


\bibliographystyle{unsrt}

\end{document}